\documentclass[journal]{IEEEtran}
\usepackage{graphicx}
\usepackage{subfigure}
\usepackage{cite}
\usepackage{bm}
\usepackage{amsmath}   % From the American Mathematical Society
\usepackage{amsthm}
\usepackage{cases}
\usepackage{amsfonts,amssymb}
\usepackage{cuted}\stripsep-3pt
\usepackage{lipsum}

\usepackage{color}
\usepackage[normalem]{ulem}
\usepackage{float}
\usepackage{stfloats}

\ifCLASSINFOpdf
\else
\fi
\begin{document}
%\title{Electromagnetic Information Theory for OAM-Based Wireless Communications}
%\title{Electromagnetic Information Theory in Analyzing Channel Capacity for OAM-Based Wireless Communications}
%\title{OAM for Wireless Communications: An Electromagnetic Information Theory Perspective}
\title{Capacity Analysis on OAM-Based Wireless Communications: An Electromagnetic Information Theory Perspective}


\author{\IEEEauthorblockN{Runyu Lyu, \textit{Student Member, IEEE}, Wenchi Cheng, \textit{Senior Member, IEEE}, Qinghe Du, \textit{Member, IEEE}, and Tony Quek, \textit{Fellow, IEEE}~}% <-this % stops an unwanted space

%\author{\IEEEauthorblockN{Runyu Lyu, Wenchi Cheng, and Muyao Wang}% <-this % stops an unwanted space
%\IEEEauthorblockA{State Key Laboratory of Integrated Services Networks, Xidian University, Xi'an, China}
%\vspace{-20pt}
\thanks{%Manuscript received March 31, 2023; revised August 15, 2023; accepted October 19, 2023. This work was supported in part by National Key R\&D Program of China under Grant 2021YFC3002102, in part by the Key R\&D Plan of Shaanxi Province under Grant 2022ZDLGY05-09, and in part by Key Area Research and Development Program of Guangdong Province under Grant 2020B0101110003. (Corresponding author: Wenchi Cheng.)

Runyu Lyu and Wenchi Cheng are with School of Telecommunications Engineering, Xidian University, Xi'an, 710071, China (e-mails: rylv@stu.xidian.edu.cn; wccheng@xidian.edu.cn).

Qinghe Du is with School of Information and Communications Engineering, Xi'an Jiaotong University, Xi'an, 710049, China (e-mail: duqinghe@mail.xjtu.edu.cn).

Tony Q.S. Quek is with Information Systems Technology and Design, Singapore University of Technology and Design, Singapore, 487372, Singapore (e-mail: tonyquek@sutd.edu.sg).
}
}

%{Shell \MakeLowercase{\textit{et al.}}: Bare Demo of IEEEtran.cls for IEEE Transactions on Magnetics Journals}



% for Transactions on Magnetics papers, we must declare the abstract and
% index terms PRIOR to the title within the \IEEEtitleabstractindextext
% IEEEtran command as these need to go into the title area created by
\IEEEtitleabstractindextext{%
%\vspace{-10pt}
\begin{abstract}
Orbital angular momentum (OAM) technology enhances the spectrum and energy efficiency of wireless communications by enabling multiplexing over different OAM modes. However, classical information theory, which relies on scalar models and far-field approximations, cannot fully capture the unique characteristics of OAM-based systems, such as their complex electromagnetic field distributions and near-field behaviors. To address these limitations, this paper analyzes OAM-based wireless communications from an electromagnetic information theory (EIT) perspective, integrating electromagnetic theory with classical information theory. EIT accounts for the physical properties of electromagnetic waves, offering advantages such as improved signal manipulation and better performance in real-world conditions. Given these benefits, EIT is more suitable for analyzing OAM-based wireless communication systems. Presenting a typical OAM model utilizing uniform circular arrays (UCAs), this paper derives the channel capacity based on the induced electric fields by using Green's function. Numerical and simulation results validate the channel capacity enhancement via exploration under EIT framework. Additionally, this paper evaluates the impact of various parameters on the channel capacity. These findings provide new insights for understanding and optimizing OAM-based wireless communications systems.
%After presenting an OAM system model utilizing uniform circular arrays (UCAs), this paper derives the channel capacity based on the induced electric fields using Green's function. Numerical and simulation results validate the enhancement of channel capacity analyzed using EIT, especially at low SNR. Additionally, this paper evaluates the impact of various parameters on the channel capacity. These findings provide new insights for understanding and optimizing OAM-based wireless communication systems.
\end{abstract}

%\vspace{-5pt}
}

\maketitle

\IEEEdisplaynontitleabstractindextext

% \fi
\IEEEpeerreviewmaketitle

\vspace{15pt}
\section{Introduction}
% OAM介绍及传统信息论分析OAM的局限性
\IEEEPARstart{O}{rbital} angular momentum (OAM)\cite{oam_light} refers to the angular momentum carried by electromagnetic waves around their propagation axis. In both optical and radio-frequency communications, OAM enhances spectrum efficiency by enabling multiplexing and demultiplexing over different OAM modes\cite{oam_low_freq_radio,oam_multiplexing,OAM_NFC_mine,OAM_Talbot} and can improve energy efficiency as well\cite{SWIPT_OAM_mine}. However, classical information theory (CIT)\cite{Shannon1948} has limitations when applied to OAM-based wireless communication systems due to the unique characteristics of OAM beams and the complexities involved in practical implementation. For example, CIT cannot fully account for the physical properties of electromagnetic waves, as it typically relies on scalar quantities, far-field approximations, discrete models, and other non-physically consistent assumptions\cite{EIT_Mag}. These assumptions are unsuitable for OAM-based wireless communication, which relies on the special structure of electromagnetic field distributions of OAM beams, emphasizes near-field characteristics, and is inherently more continuous than traditional multiple-input multiple-output (MIMO) systems\cite{oam_for_wireless_communication,oam_vs_mimo}. Additionally, while CIT provides a framework for calculating channel capacity, it can not fully capture the potential capacity gains achievable with OAM multiplexing due to the unique spatial properties of OAM beams. Studies on OAM capacity, such as \cite{oam_isItUnexploited}, use free-space channel matrices to analyze the channel capacity of OAM-based systems, reveal that communicating over the sub-channels of OAM modes can be considered a subset of MIMO and offer no additional capacity gains compared to MIMO. However, these studies do not account for the distinctive beam characteristics of OAM beams, leaving open the question of whether potential benefits have been overlooked. Therefore, a new approach that integrates electromagnetic theory and CIT is needed to study OAM-based wireless communications.

% EIT介绍及其分析OAM的适用性
Electromagnetic information theory (EIT)\cite{EIT_DoF,EIT_WirelessAntenna} is a new branch of information theory. It studies the transmission and processing of information through electromagnetic waves, integrating electromagnetic theory and CIT, which makes it capable of addressing the limitations of CIT and solving new challenges in an era where data demands are increasing\cite{6G_OAMin,Yao_5G_UAV_ResourceAllocation,zhou_UAV}. EIT offers several advantages over CIT. First, EIT, emphasizing physical layer phenomena, provides new opportunities for signal manipulation and transmission strategies that are not typically addressed in CIT\cite{EIT_Mag2}. Secondly, EIT allows for a more comprehensive analysis of channel capacity by considering the physical characteristics of electromagnetic waves, which can lead to improved data and robustness performance in real-world scenarios\cite{EIT_Mag}. Additionally, EIT supports the integration of advanced technologies, such as continuous-aperture MIMO (CAP-MIMO)\cite{CAP,CAP_PDM} and reconfigurable intelligent surfaces (RISs)\cite{Intelligent_Surfaces,RIS}. These technologies exploit the spatial and temporal degrees of freedom offered by electromagnetic waves, which are often overlooked in classical approaches. Due to its advantages over CIT, EIT is more suitable for analyzing OAM-based wireless communication systems.

\begin{figure*}[htbp]
\centering
%\vspace{-10pt}
\includegraphics[scale=0.72]{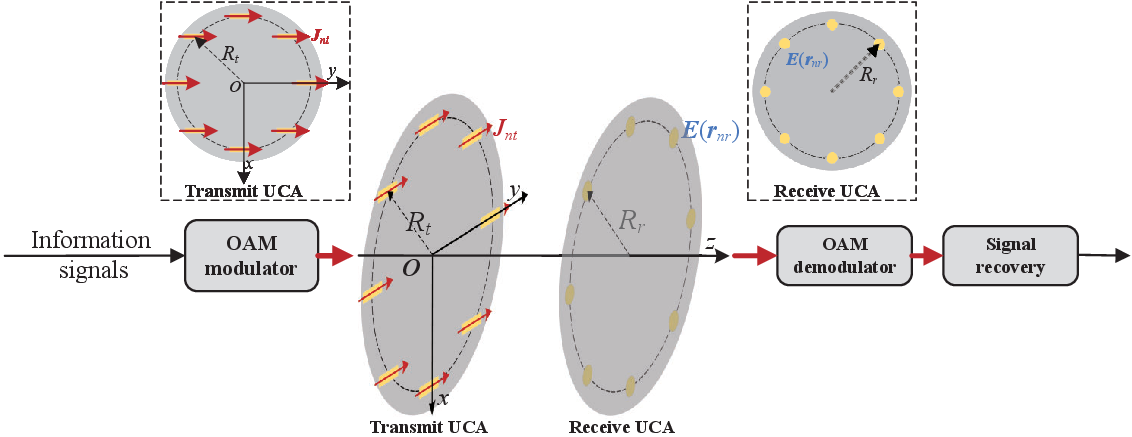}
\vspace{-5pt}
\caption{EIT based OAM wireless communication system model.} \label{fig:system_model}
\vspace{-10pt}
\end{figure*}

% 本文简介
In this paper, we analyze OAM-based wireless communications from an EIT perspective. We first propose a system model of OAM-based wireless communications based on uniform circular array (UCA) to provide an intuitive and concise analysis environment. Then, we give the induced electric field by using Green's function for continuous regions and derive the expression for the induced electric field of received OAM signal. Based on the induced electric field, we derive the channel capacity for the OAM-based wireless communication by presenting the autocorrelation and cross-correlation functions of the received OAM signals. Numerical and simulation results validate the channel capacity enhancement via exploration under EIT framework. We also analyze the impacts of different parameters on capacity, including numbers of transmitting and receiving mode, transmit and receive UCA distance, and UCA radii.

The rest of the paper is organized as follows. Section~\ref{sec:systemModel} gives the OAM-based wireless communication system model. Section~\ref{sec:OAM_EIT} derives the induced electric field of OAM signals. Channel capacity of the OAM-based wireless communication is derived in Section~\ref{sec:capacity}. Sections~\ref{sec:numericalResults} and \ref{sec:simulations} give numerical and simulation results. The conclusion is given in Section~\ref{sec:Conclusion}.

$Notation:$ Matrices and vectors are denoted by letters in bold. The notations ``$(\cdot)^*$" and ``$(\cdot)^H$" denote the conjugation and Hermitian, of a matrix or vector, respectively.

%For example, when we study continuous-aperture MIMO (CAP-MIMO)\cite{CAP,CAP_PDM}, which is a technology that extends the principles of MIMO systems to enhance both capacity and reliability by utilizing a continuous aperture for signal reception and transmission, CIT does not fully account for the physical realities of electromagnetic waves, thus unable to accurately reflect the realities of CAP-MIMO. EIT, however, studying how electromagnetic waves carry information, can provide a deeper understanding of channel capacity of CAP-MIMO by considering how the continuous aperture affects the spatial distribution of signals.

\section{System Model}\label{sec:systemModel}
%\begin{figure*}[htbp]
%\centering
%%\vspace{-10pt}
%\includegraphics[scale=0.7]{pics//system_model.eps}
%\vspace{-5pt}
%\caption{EIT based OAM wireless communication system model.} \label{fig:system_model}
%%\vspace{-5pt}
%\end{figure*}
Figure~\ref{fig:system_model} shows the OAM wireless communication system model analyzed from an EIT perspective, where $N_t$ OAM modes are used for multiplexing and $N_r$ OAM modes for demultiplexing. A UCA, consisting of $N_t$ linear sources, is employed as the transmit source. The length of each linear source is set to half the wavelength and aligned along the $y$-axis. The current density at the $n_t$th transmit source is denoted by $\boldsymbol{\mathrm J}_{n_t}\in \mathbb{C}^{3\times 1}$, which is also aligned along the $y$-axis. The receiver is also a UCA, but consisting of $N_r$ point antennas. The induced electric field at the $n_r$th receive point is represented as $\boldsymbol{\mathrm E}_{\boldsymbol{\mathrm r}_{n_r}}\in \mathbb{C}^{3\times 1}$, where $\boldsymbol{\mathrm r}_{n_r}\in \mathbb{R}^{3\times 1}$ denotes the coordinates of the $n_r$th receive point. The radii of the transmit and receive UCAs are denoted by $R_t$ and $R_r$, respectively. UCAs are selected due to their ability to digitally generate OAM beams with a uniform radius and common center\cite{oam_generation_by_uca_2017}. However, it should be noted that other antenna types, such as spiral phase plates or metasurfaces with different geometries, can also be used to generate OAM beams. As depicted in Fig.~\ref{fig:system_model}, the multi-stream RF information signals are first modulated, allocated power across different OAM modes, and then transmitted via the transmit UCA. The induced electric field is generated at the receive UCA, where the signals are demodulated to extract the OAM-modulated information.

%\section{OAM Wireless Communications Analyzed through EIT}\label{sec:OAM_EIT}
\section{Analysis on OAM Wireless Communications from EIT Perspective}\label{sec:OAM_EIT}
In this section, we first present the induced electric field by using Green's function for continuous regions. Following this, we derive the expression for the induced electric field of OAM-based system as well as the received OAM signal. Finally, we provide some figures of the induced electric field for different OAM modes.

\subsection{Wireless Communication of Continuous Regions}
In EIT, Green's function can be used to model the wireless communication between two continuous regions. Let $\boldsymbol{\mathrm J}(\boldsymbol{\mathrm s})\in \mathbb{C}^{3\times 1}$ and $\boldsymbol{\mathrm E}(\boldsymbol{\mathrm r})\in \mathbb{C}^{3\times 1}$ denote the current density at the source and the induced electric field at the destination respectively, where $\boldsymbol{\mathrm s}\in \mathbb{R}^{3\times 1}$ represents the coordinate of the source and $\boldsymbol{\mathrm r}\in \mathbb{R}^{3\times 1}$ represents the coordinate of the field observer. The induced electric filed can be derived through the current density by using Green's function as follows:
\begin{align}
\boldsymbol{\mathrm E}(\boldsymbol{\mathrm r}) = \int_{V_s}\boldsymbol{\mathrm G}(\boldsymbol{\mathrm r},\boldsymbol{\mathrm s})\boldsymbol{\mathrm J}(\boldsymbol{\mathrm s}){\mathrm d}\boldsymbol{\mathrm s},\ \boldsymbol{\mathrm r}\in V_r,
\label{eq:EGJ}
\end{align}
where $\boldsymbol{\mathrm G}(\boldsymbol{\mathrm r},\boldsymbol{\mathrm s})\in \mathbb{C}^{3\times 3}$ denotes the Green's function, $V_s$ denotes the source space, and $V_r$ denotes the received space. In unbounded, homogeneous mediums at a fixed frequency point, the Green's function can be simplified as follows\cite{EIT_RandomFieldsMutualInformation}:
\begin{align}
\boldsymbol{\mathrm G}(\boldsymbol{\mathrm r},\boldsymbol{\mathrm s}) &= \frac{jk_0Z_0}{4\pi\Vert\boldsymbol{\mathrm p}\Vert}\left(\boldsymbol{\mathrm I}_3+\frac{\nabla_{\boldsymbol{\mathrm r}}\nabla_{\boldsymbol{\mathrm r}}^H}{k_0^2}\right)e^{jk_0\Vert\boldsymbol{\mathrm p}\Vert}\nonumber\\
&\quad\approx \frac{jk_0Z_0e^{jk_0\Vert\boldsymbol{\mathrm p}\Vert}}{4\pi\Vert\boldsymbol{\mathrm p}\Vert}\Bigg[(\boldsymbol{\mathrm I}_3-\hat{\boldsymbol{\mathrm p}}\hat{\boldsymbol{\mathrm p}}^H) \nonumber\\
&\quad\quad\quad+ \frac{j\lambda}{2\pi\Vert\boldsymbol{\mathrm p}\Vert}(\boldsymbol{\mathrm I}_3-3\hat{\boldsymbol{\mathrm p}}\hat{\boldsymbol{\mathrm p}}^H) \nonumber\\
&\quad\quad\quad\quad- \left(\frac{\lambda}{2\pi\Vert\boldsymbol{\mathrm p}\Vert}\right)^2(\boldsymbol{\mathrm I}_3-3\hat{\boldsymbol{\mathrm p}}\hat{\boldsymbol{\mathrm p}}^H)\Bigg],
\label{eq:GreenFunc}
\end{align}
where $j$ denotes the imaginary unit, $k_0$ represents the wave number, $Z_0$ denotes the free-space intrinsic impedance, $\boldsymbol{\mathrm p}=\boldsymbol{\mathrm r}-\boldsymbol{\mathrm s}$ denotes the distance vector, $\Vert\boldsymbol{\mathrm p}\Vert$ does the $l^2$-norm of $\boldsymbol{\mathrm p}$, $\hat{\boldsymbol{\mathrm p}}=\boldsymbol{\mathrm p}/\Vert\boldsymbol{\mathrm p}\Vert$ denotes the distance unit vector, $\boldsymbol{\mathrm I}_3$ denotes an identity matrix of size $3\times3$, and $\lambda$ represents the wave length. %The received electric field, denoted by $\boldsymbol{\mathrm Y}(\boldsymbol{\mathrm r})\in \mathbb{C}^{3\times 1}$, can be further given as $\boldsymbol{\mathrm Y}(\boldsymbol{\mathrm r}) = \boldsymbol{\mathrm E}(\boldsymbol{\mathrm r}) + \boldsymbol{\mathrm N}(\boldsymbol{\mathrm r})$, where $\boldsymbol{\mathrm N}(\boldsymbol{\mathrm r})\in \mathbb{C}^{3\times 1}$ denotes the noise in the observed field.

%\subsection{OAM Wireless Communication Analyzed through EIT}
\subsection{Analysis on OAM Wireless Communications Based on EIT}
Considering the OAM wireless communications as shown in Fig.~\ref{fig:system_model}, the current densities of the sources are along the $y$-axis and are given by multiplying the multi-stream RF information signal, denoted by $\boldsymbol{\mathrm x}\in \mathbb{C}^{N_t\times 1}$, with a normalized IDFT matrix, denoted by $\boldsymbol{\mathrm W}\in \mathbb{C}^{N_t\times Nt}$ with the $n_1$th row and the $n_2$th column element given by ${W}_{n_1,n_2}\hspace{-0.1cm}=\hspace{-0.1cm}\frac{1}{\sqrt N_t}{\rm exp}[j{2\pi (n_1\hspace{-0.1cm}-\hspace{-0.1cm}1)(n_2\hspace{-0.1cm}-\hspace{-0.1cm}1)}/{N_t}]$. Thus, the obtained current density of the $n_t$th transmit source can be given as follows:
\begin{align}
\boldsymbol{\mathrm J}_{n_t} \hspace{-0.1cm}=\hspace{-0.1cm} \boldsymbol{\mathrm J}_{n_t} \hspace{-0.1cm}\cdot\hspace{-0.05cm} \hat{\boldsymbol{\mathrm e}}_y \hspace{-0.1cm}=\hspace{-0.1cm} J^y_{n_t} \hspace{-0.1cm}=\hspace{-0.1cm} \frac{1}{\sqrt{N_t}}\sum_{m=1}^{N_t}x(m)e^{\frac{j2\pi(m-1)(n_t-1)}{N_t}},
\label{eq:J}
\end{align}
where $\hat{\boldsymbol{\mathrm e}}_y = [0,1,0]^T$ denotes the unit vector along $y$ coordinate, $J^y_{n_t}$ represents the $y$ component for the current density of the $n_t$th transmit source, and $x(m)$ denotes the $m$th element of $\boldsymbol{\mathrm x}$. Thus, Eq.~\eqref{eq:EGJ} can be rewritten as follows:
\begin{align}
\boldsymbol{\mathrm E}(\boldsymbol{\mathrm r}) &= \int_{V_{s,n_t}}\boldsymbol{\mathrm G}(\boldsymbol{\mathrm r},\boldsymbol{\mathrm s})\boldsymbol{\mathrm J}(\boldsymbol{\mathrm s}){\mathrm d}\boldsymbol{\mathrm s}\nonumber\\
&\quad=\hspace{-0.1cm}\sum_{n_t = 1}^{N_t}\hspace{-0.1cm}\int_{R_t\sin(\frac{2\pi n_t}{N_t})-\frac{L}{2}}^{R_t\sin(\frac{2\pi n_t}{N_t})+\frac{L}{2}}\hspace{-0.1cm}\boldsymbol{\mathrm G}(\boldsymbol{\mathrm r},\boldsymbol{\mathrm s})J^y_{n_t}{\mathrm d}s^y\nonumber\\
%&\quad\quad=\sum_{n_t = 1}^{N_t}\int_{y_{n_t}-\frac{L}{2}}^{y_{n_t}+\frac{L}{2}}
%\begin{bmatrix}G^{xy}(\boldsymbol{\mathrm r},\boldsymbol{\mathrm s}) \\ G^{yy}(\boldsymbol{\mathrm r},\boldsymbol{\mathrm s}) \\ G^{zy}(\boldsymbol{\mathrm r},\boldsymbol{\mathrm s})\end{bmatrix}
%J^y_{n_t}{\mathrm d} s^y\nonumber\\
&\quad\quad=\sum_{n_t = 1}^{N_t}
J^y_{n_t}\int_{y_{n_t}-\frac{L}{2}}^{y_{n_t}+\frac{L}{2}}
\begin{bmatrix}G^{xy}(\boldsymbol{\mathrm r},\boldsymbol{\mathrm s}) \\ G^{yy}(\boldsymbol{\mathrm r},\boldsymbol{\mathrm s}) \\ G^{zy}(\boldsymbol{\mathrm r},\boldsymbol{\mathrm s})\end{bmatrix}{\mathrm d} s^y,\ \boldsymbol{\mathrm s}\in V_{s,n_t},
\label{eq:EGJ2}
\end{align}
where $V_{s,n_t}$ denotes the source space of the $n_t$th antenna, $s^y$ denotes the $y$ coordinate of $\boldsymbol{\mathrm s}$, $y_{n_t} = R_t\sin(\frac{2\pi n_t}{N_t})$ denotes the $y$ coordinate of the $n_t$th transmit source, $G^{xy}(\boldsymbol{\mathrm r},\boldsymbol{\mathrm s})$, $G^{yy}(\boldsymbol{\mathrm r},\boldsymbol{\mathrm s})$, and $G^{zy}(\boldsymbol{\mathrm r},\boldsymbol{\mathrm s})$ denote the first, second, and third elements in the second column of $\boldsymbol{\mathrm G}(\boldsymbol{\mathrm r},\boldsymbol{\mathrm s})$ respectively. When using high-resolution discrete lens arrays to form a quasi-continuous aperture phased UCA, Eq.~\eqref{eq:EGJ2} can be rewritten into a discrete form as follows:
\begin{align}
\boldsymbol{\mathrm E}(\boldsymbol{\mathrm r}) = \frac{1}{N_l}\sum_{n_t = 1}^{N_t}\sum_{n_l = 1}^{N_l}\begin{bmatrix}G^{xy}(\boldsymbol{\mathrm r},\boldsymbol{\mathrm s}_{n_t,n_l}) \\ G^{yy}(\boldsymbol{\mathrm r},\boldsymbol{\mathrm s}_{n_t,n_l}) \\ G^{zy}(\boldsymbol{\mathrm r},\boldsymbol{\mathrm s}_{n_t,n_l})\end{bmatrix}
J^y_{n_t},
\label{eq:EGJ_D}
\end{align}
where $\boldsymbol{\mathrm s}_{n_t,n_l}$ denotes the coordinate for the $n_l$th small feed of the $n_t$th transmit source in the UCA and $s^y_{n_t,n_l}$ is the $y$ coordinate of $\boldsymbol{\mathrm s}_{n_t,n_l}$. The value of $\boldsymbol{\mathrm s}_{n_t,n_l}$ is given as follows:
\begin{align}
\boldsymbol{\mathrm s}_{n_t,n_l} =
\begin{bmatrix}
R_t\cos(\frac{2\pi n_t}{N_t})\\
R_t\sin(\frac{2\pi n_t}{N_t}) + \frac{n_l-1}{N_l-1}L -\frac{L}{2}\\
0
\end{bmatrix}.
\label{eq:sntnl}
\end{align}

As shown in Fig.~\ref{fig:system_model}, the received signal is given as the received induced electric field multiplied with a normalized DFT matrix. Therefore, the $l$th received OAM signal, denoted by $\boldsymbol{\mathrm y}_l$, is given by performing DFT on the induced electric field at receive points as follows:
\begin{align}
\boldsymbol{\mathrm y}_l = \frac{1}{\sqrt{N_r}}\sum_{n_r=1}^{N_r}\boldsymbol{\mathrm E}(\boldsymbol{\mathrm r}_{n_r}) e^{\frac{j2\pi l(n_r-1)}{N_r}} + \boldsymbol{\mathrm n},
\label{eq:Y}
\end{align}
where $\boldsymbol{\mathrm n}$ denotes the additive noise.

\begin{figure}[htbp]
\centering
%\vspace{-10pt}
\includegraphics[scale=0.52]{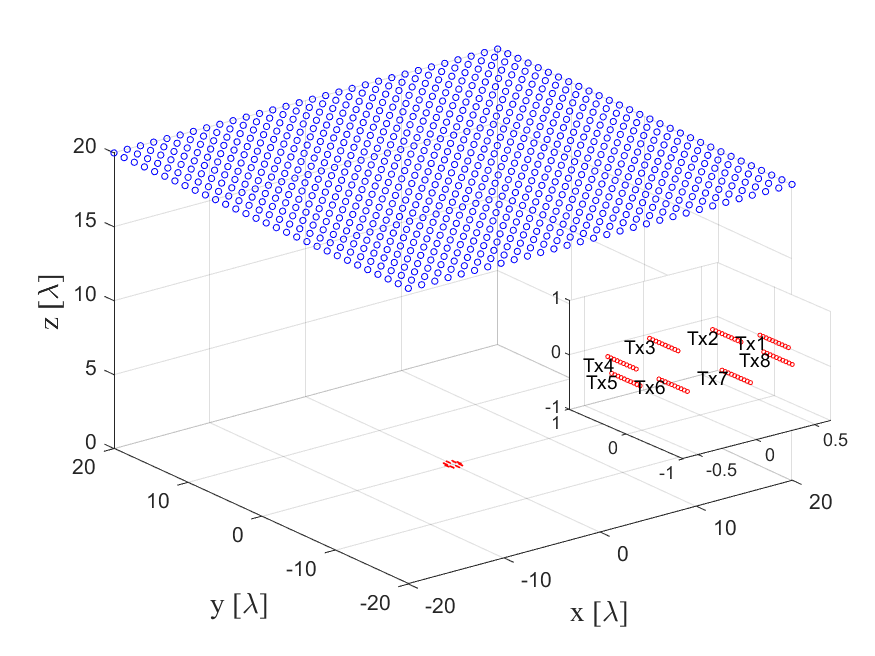}
%\vspace{-5pt}
\caption{Transmit UCA and observer schematic.} \label{fig:TxObserver}
%\vspace{-5pt}
\end{figure}
\begin{figure}[!h]
\centering
%\vspace{-15pt}
\subfigure[OAM-mode $-1$.]{
\begin{minipage}{0.45\linewidth}
\centering
\includegraphics[scale=0.32]{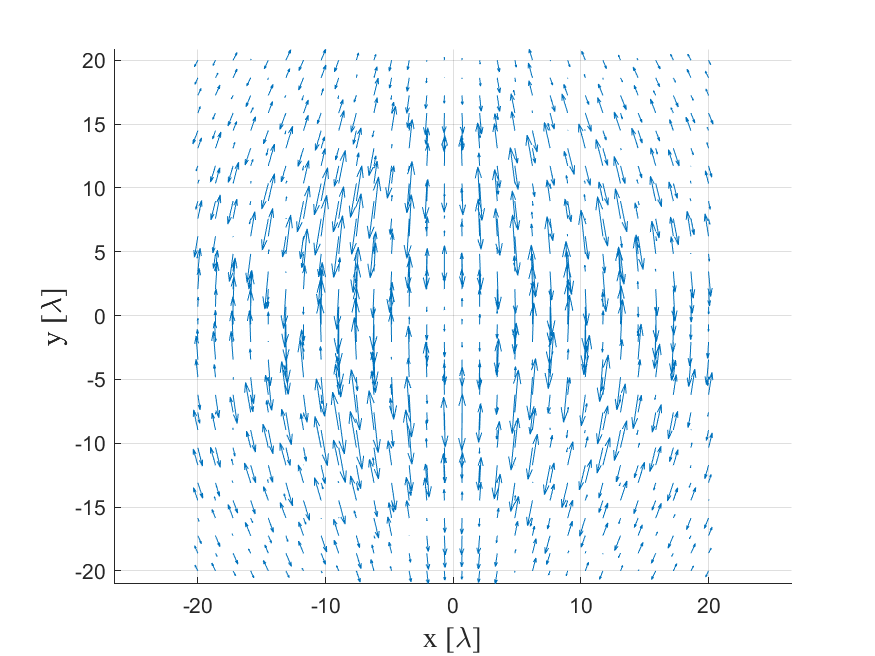}
%\vspace{-30pt}
\end{minipage}
}
\subfigure[OAM-mode $0$.]{
\begin{minipage}{0.45\linewidth}
\centering
\includegraphics[scale=0.32]{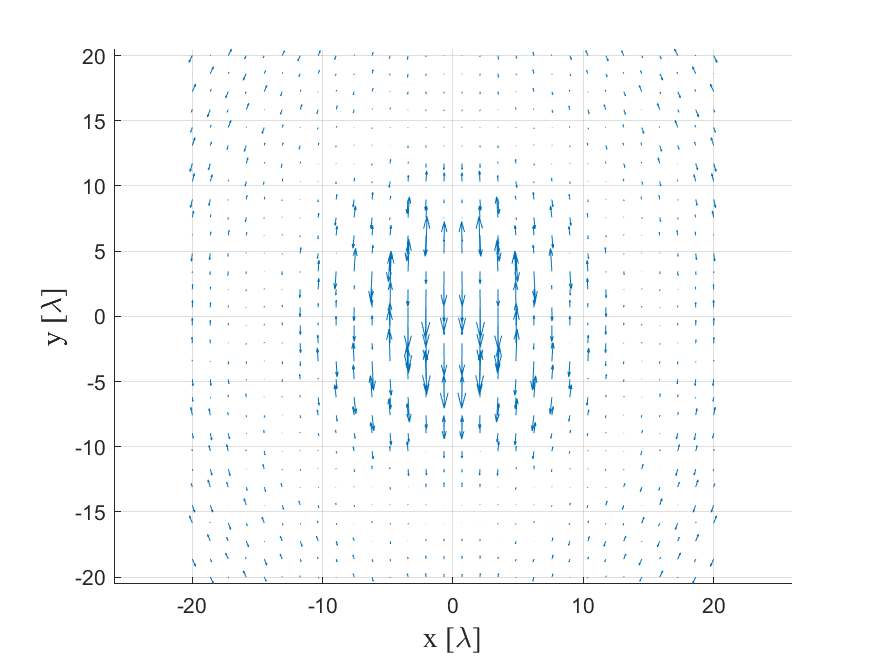}
%\vspace{-30pt}
\end{minipage}
}
\subfigure[OAM-mode $1$.]{
\begin{minipage}{0.45\linewidth}
\centering
\includegraphics[scale=0.32]{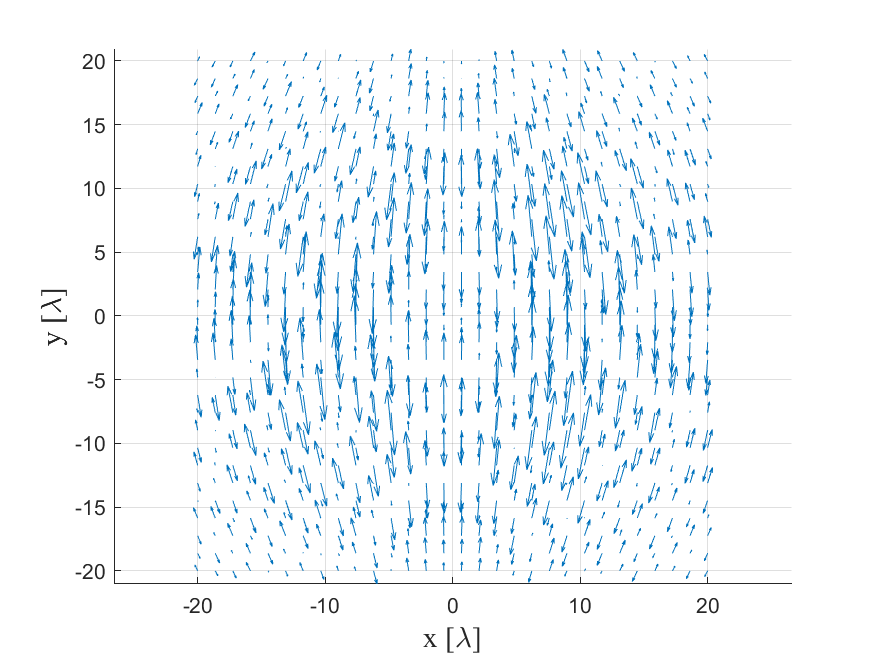}
%\vspace{-30pt}
\end{minipage}
}
\subfigure[OAM-mode $2$.]{
\begin{minipage}{0.45\linewidth}
\centering
\includegraphics[scale=0.32]{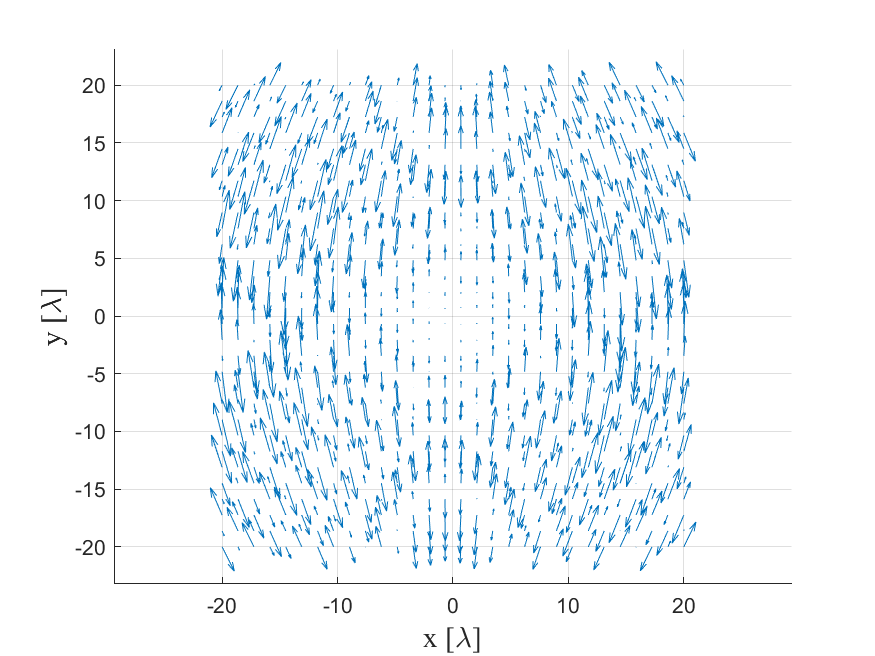}
%\vspace{-30pt}
\end{minipage}
}
\centering
%\vspace{-10pt}
\caption{Electric fields of OAM-mode $-1$, $0$, $1$, and $2$.} \label{fig:eFiled}
%\vspace{-15pt}
\end{figure}
To better illustrate the induced electric field of the OAM signals, we deploy a grid of $100\times 100$ virtual receive points on a plane located $20\lambda$ ($1034.5$ mm) away from the transmit UCA as shown in Fig.~\ref{fig:TxObserver}. The frequency is set to $5.8$ GHz. The width of the observer plane is also set to $20\lambda$. The number of transmit sources is set to $8$, each consisting of $30$ small feeds. The current density is normalized by dividing it by $\sqrt{240}$ for each feed. The length of each transmit source is set to $0.5\lambda$ ($25.9$ mm) and the radius of the transmit UCA is set to $2\lambda/\pi$ ($32.9$ mm). Based on the above settings, Fig.~\ref{fig:eFiled} presents the induced electric fields for OAM modes $-1$ to $2$ with the phase set to $0$ deg. As shown in the figure, the induced electric fields are primarily oriented along the $y$-axis, same to the current densities at the source. Fig.~\ref{fig:eFiled} also highlights the hollow divergence phenomenon characteristic of OAM transmissions.

%Additionally, we observe that the current density directions for OAM modes $-1$ and $1$ are opposite.

\begin{figure}[!h]
\centering
%\vspace{-15pt}
\subfigure[$x$ component phase.]{
\begin{minipage}{0.45\linewidth}
\centering
\includegraphics[scale=0.28]{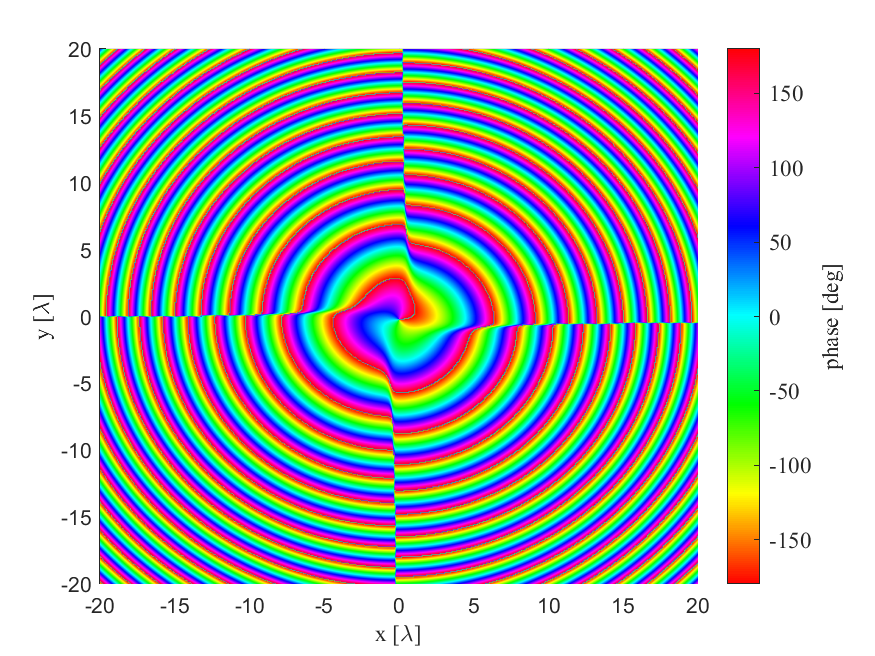}
%\vspace{-30pt}
\end{minipage}
}
\subfigure[$x$ component power.]{
\begin{minipage}{0.45\linewidth}
\centering
\includegraphics[scale=0.28]{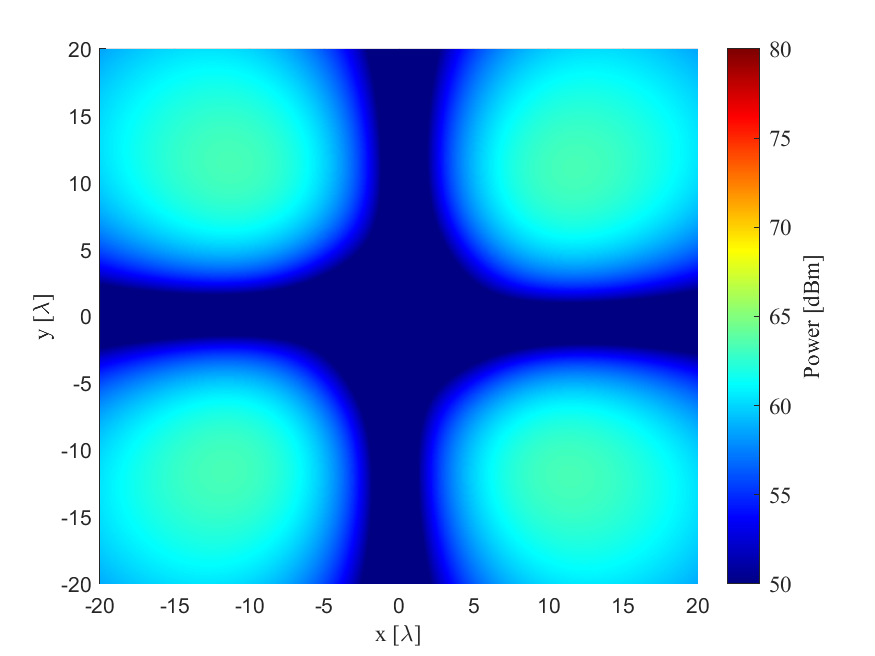}
%\vspace{-30pt}
\end{minipage}
}
\subfigure[$y$ component phase.]{
\begin{minipage}{0.45\linewidth}
\centering
\includegraphics[scale=0.28]{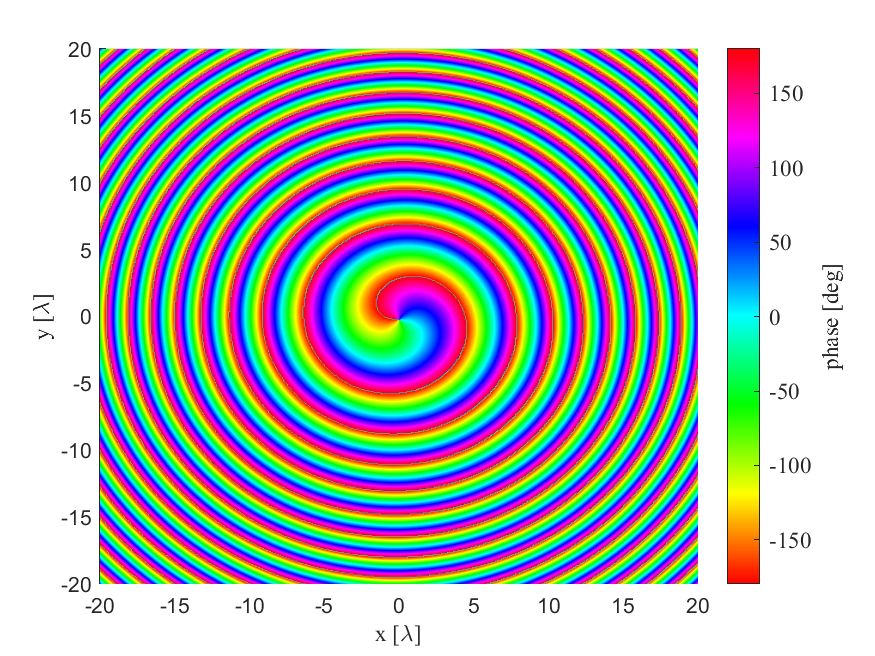}
%\vspace{-30pt}
\end{minipage}
}
\subfigure[$y$ component power.]{
\begin{minipage}{0.45\linewidth}
\centering
\includegraphics[scale=0.28]{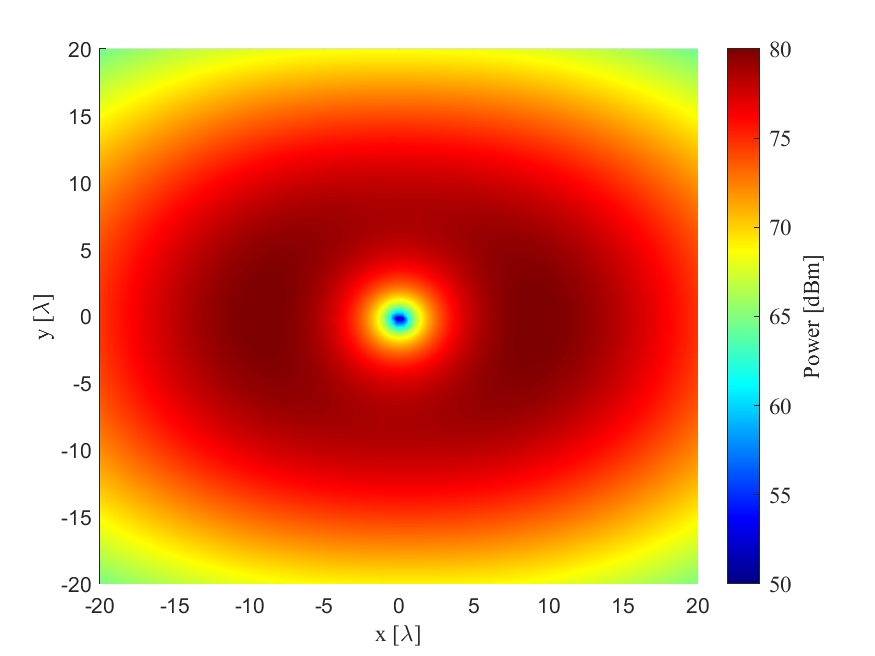}
%\vspace{-30pt}
\end{minipage}
}
\subfigure[$z$ component phase.]{
\begin{minipage}{0.45\linewidth}
\centering
\includegraphics[scale=0.28]{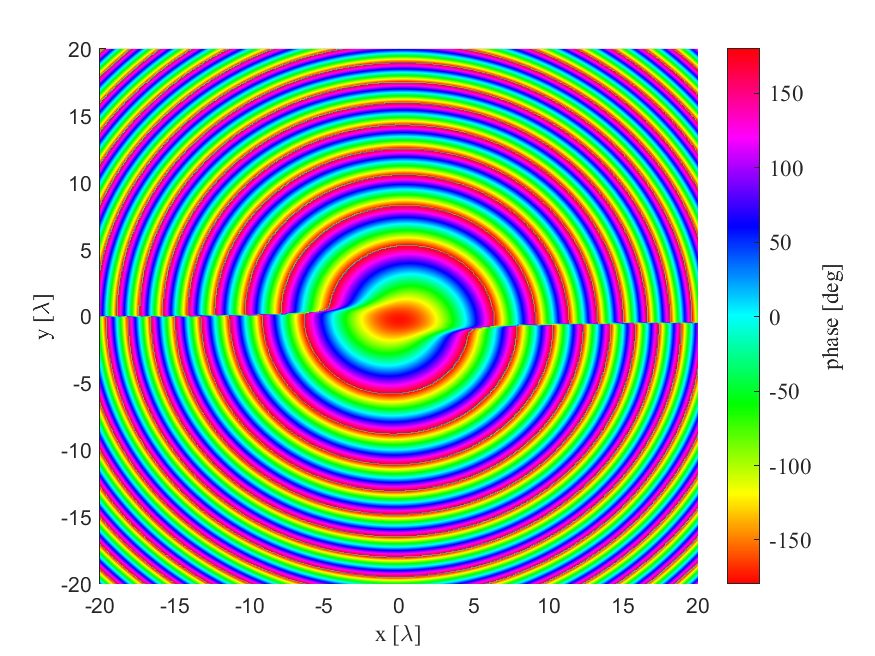}
%\vspace{-30pt}
\end{minipage}
}
\subfigure[$z$ component power.]{
\begin{minipage}{0.45\linewidth}
\centering
\includegraphics[scale=0.28]{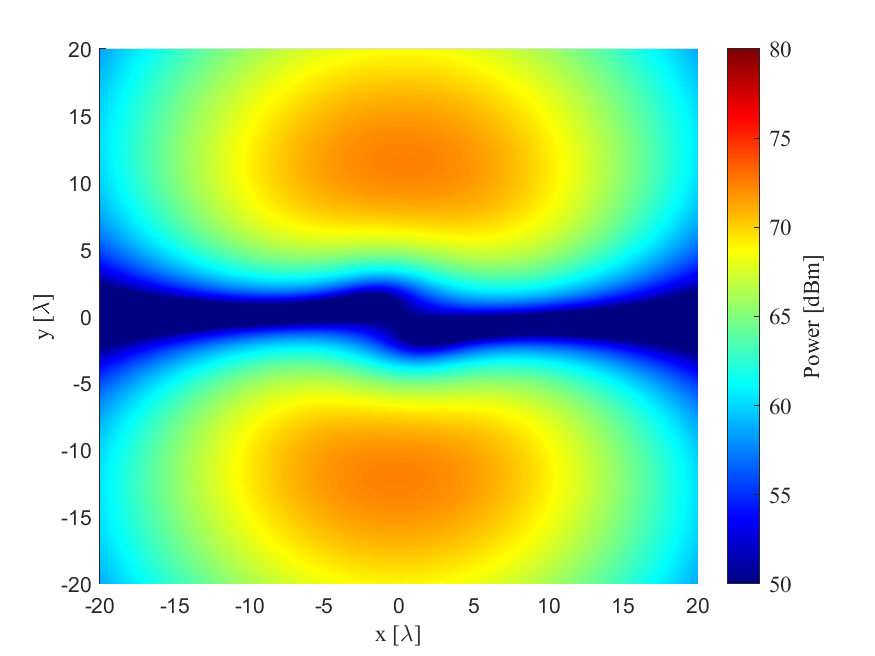}
%\vspace{-30pt}
\end{minipage}
}
\centering
%\vspace{-10pt}
\caption{$x$, $y$, and $z$ components of power and phase for OAM-mode $1$.} \label{fig:powerPhaseMode1}
%\vspace{-15pt}
\end{figure}
We also plot the $x$, $y$, and $z$ components of the power and phase for OAM mode $1$ in Fig.~\ref{fig:powerPhaseMode1} based on the same setting in Fig.~\ref{fig:TxObserver}. As shown in this figure, the power of the $y$-component is significantly higher than that of the $x$ and $z$-components, indicating that the primary direction of the induced electric field aligns with the current density direction of the source. The typical spiral phase structure of OAM beams is also observed in the $y$-direction. Additionally, Fig.~\ref{fig:powerPhaseMode1} highlights the hollow divergence of OAM beams.

\section{Channel Capacity Derivation under EIT Framework}\label{sec:capacity}
In this section, we derive the channel capacity for the OAM-based wireless communications from an EIT perspective. We begin by presenting the autocorrelation and cross-correlation functions of the received OAM signals. Based on the cross-correlation matrix of the received OAM signals, we then calculate the channel capacity.

\subsection{Autocorrelation and Cross-correlation Functions of Received OAM Signals}
The cross-correlation function between the $n_t$th and $n'_t$th $y$ component current density, denoted by ${\boldsymbol{\mathrm R}}_{J^y_{n_t},J^y_{n'_t}}\hspace{-0.1cm}=\hspace{-0.1cm}\mathbb{E}\left\{J^y_{n_t}(J^y_{n'_t})^*\right\}$, can be give as follows:
\begin{align}
{\boldsymbol{\mathrm R}}_{J^y_{n_t},J^y_{n'_t}}\hspace{-0.1cm}&=\hspace{-0.1cm}\mathbb{E}\Bigg\{\frac{1}{\sqrt{N_t}}\sum_{m=1}^{N_t}x(m)e^{\frac{j2\pi(m-1)(n_t-1)}{N_t}}\nonumber\\
&\hspace{1.5cm}\frac{1}{\sqrt{N_t}}\sum_{m'=1}^{N_t}x^*(m')e^{-\frac{j2\pi(m'-1)(n'_t-1)}{N_t}}\Bigg\}\nonumber\\
&\quad=\hspace{-0.1cm}\frac{1}{N_t}\hspace{-0.1cm}\sum_{m=1}^{N_t}\sum_{m'=1}^{N_t}\hspace{-0.1cm}\mathbb{E}\left\{x(m)x^*(m')\right\}\nonumber\\
&\hspace{1.5cm}\quad\quad\quad\quad e^{\frac{j2\pi(mn_t\hspace{-0.04cm}-\hspace{-0.04cm}m'n'_t\hspace{-0.04cm}-\hspace{-0.04cm}m\hspace{-0.04cm}+\hspace{-0.04cm}m'\hspace{-0.04cm}-\hspace{-0.04cm}n_t\hspace{-0.04cm}+\hspace{-0.04cm}n'_t)}{N_t}}.
\label{eq:R_J}
\end{align}
For normalized input signals independent from each other, we have $\mathbb{E}\left\{\boldsymbol{\mathrm x}\boldsymbol{\mathrm x}^H\right\}\hspace{-0.1cm}=\hspace{-0.1cm}\boldsymbol{\mathrm I}_{N_t}$. Therefore, Eq.~\eqref{eq:R_J} can be rewritten as follows:
\begin{align}
{\boldsymbol{\mathrm R}}_{J^y_{n_t},J^y_{n'_t}}\hspace{-0.1cm}=\hspace{-0.1cm}\frac{1}{N_t}\hspace{-0.1cm}\sum_{m=1}^{N_t}e^{\frac{j2\pi(m-1)(n_t-n'_t)}{N_t}}
=\begin{cases}
0,\ {\rm n_t\neq n'_t}\\
1,\ {\rm n_t=n'_t}.
\end{cases}
\label{eq:R_J2}
\end{align}
Base on Eq.~\eqref{eq:EGJ2}, the autocorrelation function of induced electric field, denoted by ${\boldsymbol{\mathrm R}}_{\boldsymbol{\mathrm E}}(\boldsymbol{\mathrm r},\boldsymbol{\mathrm r}')\hspace{-0.1cm}=\hspace{-0.1cm}\mathbb{E}\left\{\boldsymbol{\mathrm E}(\boldsymbol{\mathrm r})\boldsymbol{\mathrm E}^H(\boldsymbol{\mathrm r}')\right\}$, can be derived as follows:
\begin{align}
&{\boldsymbol{\mathrm R}}_{\boldsymbol{\mathrm E}}(\boldsymbol{\mathrm r},\boldsymbol{\mathrm r}')\nonumber\\
&\quad=\hspace{-0.1cm} \mathbb{E}\Bigg\{\hspace{-0.1cm}\left[\sum_{n_t = 1}^{N_t}
\hspace{-0.1cm}J^y_{n_t}\hspace{-0.1cm}\int_{y_{n_t}-\frac{L}{2}}^{y_{n_t}+\frac{L}{2}}\hspace{-0.1cm}
\begin{bmatrix}G^{xy}(\boldsymbol{\mathrm r},\boldsymbol{\mathrm s}) \\ G^{yy}(\boldsymbol{\mathrm r},\boldsymbol{\mathrm s}) \\ G^{zy}(\boldsymbol{\mathrm r},\boldsymbol{\mathrm s})\end{bmatrix}\hspace{-0.1cm}{\mathrm d} s^y\right]\nonumber\\
&\hspace{2cm}\left[\sum_{n'_t = 1}^{N_t}
\hspace{-0.1cm}J^y_{n'_t}\hspace{-0.1cm}\int_{y_{n'_t}-\frac{L}{2}}^{y_{n'_t}+\frac{L}{2}}
\hspace{-0.1cm}\begin{bmatrix}G^{xy}(\boldsymbol{\mathrm r}',\boldsymbol{\mathrm s}') \\ G^{yy}(\boldsymbol{\mathrm r}',\boldsymbol{\mathrm s}') \\ G^{zy}(\boldsymbol{\mathrm r}',\boldsymbol{\mathrm s}')\end{bmatrix}\hspace{-0.1cm}{\mathrm d} s'^y\right]^H\hspace{-0.1cm}\Bigg\}\nonumber\\
&\quad\quad= \hspace{-0.1cm}\sum_{n_t = 1}^{N_t}\sum_{n'_t = 1}^{N_t}{\boldsymbol{\mathrm R}}_{J^y_{n_t},J^y_{n'_t}}\hspace{-0.1cm}\int_{y_{n_t}-\frac{L}{2}}^{y_{n_t}+\frac{L}{2}}\hspace{-0.1cm}\int_{y_{n'_t}-\frac{L}{2}}^{y_{n'_t}+\frac{L}{2}} \nonumber\\
&\hspace{2cm}\quad\begin{bmatrix}G^{xy}(\boldsymbol{\mathrm r},\boldsymbol{\mathrm s}) \\ G^{yy}(\boldsymbol{\mathrm r},\boldsymbol{\mathrm s}) \\ G^{zy}(\boldsymbol{\mathrm r},\boldsymbol{\mathrm s})\end{bmatrix}\hspace{-0.1cm}\begin{bmatrix}G^{xy}(\boldsymbol{\mathrm r}',\boldsymbol{\mathrm s}') \\ G^{yy}(\boldsymbol{\mathrm r}',\boldsymbol{\mathrm s}') \\ G^{zy}(\boldsymbol{\mathrm r}',\boldsymbol{\mathrm s}')\end{bmatrix}^H\hspace{-0.3cm}{\mathrm d}s^y{\mathrm d}s'^y,\nonumber\\
&\hspace{3.8cm}\boldsymbol{\mathrm r},\boldsymbol{\mathrm r}'\in V_r,\boldsymbol{\mathrm s}\in V_{s,n_t},\boldsymbol{\mathrm s}'\in V_{s,n'_t}.
\label{eq:E_autocorr}
\end{align}
Substitute Eq.~\eqref{eq:R_J2} into Eq.~\eqref{eq:E_autocorr}, the autocorrelation function of induced electric field can be given as follows:
\begin{small}
\begin{align}
{\boldsymbol{\mathrm R}}_{\boldsymbol{\mathrm E}}(\boldsymbol{\mathrm r},\boldsymbol{\mathrm r}')\hspace{-0.1cm}=\hspace{-0.1cm}\hspace{-0.1cm}\sum_{n_t = 1}^{N_t}\hspace{-0.1cm}\int_{y_{n_t}-\frac{L}{2}}^{y_{n_t}+\frac{L}{2}}\hspace{-0.15cm}\int_{y_{n_t}-\frac{L}{2}}^{y_{n_t}+\frac{L}{2}}
\hspace{-0.1cm}\begin{bmatrix}G^{xy}(\boldsymbol{\mathrm r},\boldsymbol{\mathrm s}) \\ G^{yy}(\boldsymbol{\mathrm r},\boldsymbol{\mathrm s}) \\ G^{zy}(\boldsymbol{\mathrm r},\boldsymbol{\mathrm s})\end{bmatrix}\hspace{-0.1cm}\begin{bmatrix}G^{xy}(\boldsymbol{\mathrm r}',\boldsymbol{\mathrm s}') \\ G^{yy}(\boldsymbol{\mathrm r}',\boldsymbol{\mathrm s}') \\ G^{zy}(\boldsymbol{\mathrm r}',\boldsymbol{\mathrm s}')\end{bmatrix}^H\hspace{-0.3cm}{\mathrm d}s^y{\mathrm d}s'^y.
\label{eq:E_autocorr2}
\end{align}
\end{small}
\hspace{-0.18cm}
The discrete form of Eq.~\eqref{eq:E_autocorr2} can be given as follows:
\begin{small}
\begin{align}
{\boldsymbol{\mathrm R}}_{\boldsymbol{\mathrm E}}(\boldsymbol{\mathrm r},\boldsymbol{\mathrm r}') &\hspace{-0.1cm}=\hspace{-0.1cm}\frac{1}{N^2_l}\hspace{-0.1cm}\sum_{n_t = 1}^{N_t}\sum_{n_l = 1}^{N_l}\sum_{n'_l = 1}^{N_l}\begin{bmatrix}G^{xy}(\boldsymbol{\mathrm r},\boldsymbol{\mathrm s}_{n_t,n_l}) \\ G^{yy}(\boldsymbol{\mathrm r},\boldsymbol{\mathrm s}_{n_t,n_l}) \\ G^{zy}(\boldsymbol{\mathrm r},\boldsymbol{\mathrm s}_{n_t,n_l})\end{bmatrix}\hspace{-0.1cm}\begin{bmatrix}G^{xy}(\boldsymbol{\mathrm r},\boldsymbol{\mathrm s}_{n_t,n'_l}) \\ G^{yy}(\boldsymbol{\mathrm r},\boldsymbol{\mathrm s}_{n_t,n'_l}) \\ G^{zy}(\boldsymbol{\mathrm r},\boldsymbol{\mathrm s}_{n_t,n'_l})\end{bmatrix}^H.
\label{eq:E_autocorr_D}
\end{align}
\end{small}

Based on Eq.~\eqref{eq:Y} and let the noise independent of the signals, the cross-correlation function between the received signals of the $l$th and $l'$th OAM modes, denoted by ${\boldsymbol{\mathrm R}}_{\boldsymbol{\mathrm y}}(l,l')\hspace{-0.1cm}=\hspace{-0.1cm}\mathbb{E}\left\{\boldsymbol{\mathrm y}_l\boldsymbol{\mathrm y}^H_{l'}\right\}$, can be derived as follows:
\begin{align}
{\boldsymbol{\mathrm R}}_{\boldsymbol{\mathrm y}}(l,l')\hspace{-0.1cm}&=\hspace{-0.1cm}\mathbb{E}\Bigg\{\frac{1}{\sqrt{N_r}}\sum_{n_r=1}^{N_r}\boldsymbol{\mathrm E}(\boldsymbol{\mathrm r}_{n_r}) e^{\frac{j2\pi l(n_r-1)}{N_r}} + \boldsymbol{\mathrm n},\nonumber\\
&\hspace{1.2cm}\left(\frac{1}{\sqrt{N_r}}\sum_{n'_r=1}^{N_r}\boldsymbol{\mathrm E}(\boldsymbol{\mathrm r}_{n'_r}) e^{\frac{j2\pi l'(n'_r-1)}{N_r}} + \boldsymbol{\mathrm n}'\right)^H\Bigg\}\nonumber\\
&\quad=\hspace{-0.1cm}\frac{1}{N_r}\sum_{n_r=1}^{N_r}\sum_{n'_r=1}^{N_r}{\boldsymbol{\mathrm R}}_{\boldsymbol{\mathrm E}}(\boldsymbol{\mathrm r}_{n_r},\boldsymbol{\mathrm r}_{n'_r}) e^{\frac{j2\pi \hspace{-0.04cm}\left[l(n_r\hspace{-0.04cm}-\hspace{-0.04cm}1)\hspace{-0.04cm}-\hspace{-0.04cm}l'(n'_r\hspace{-0.04cm}-\hspace{-0.04cm}1)\right]}{N_r}}.
\label{eq:R_y}
\end{align}
Substitute Eq.~\eqref{eq:E_autocorr2} into \eqref{eq:R_y}, the cross-correlation function between the received signals of different OAM modes can be rewritten as follows:
\begin{small}
\begin{align}
{\boldsymbol{\mathrm R}}_{\boldsymbol{\mathrm y}}(l,l')\hspace{-0.1cm}&=\hspace{-0.1cm}\frac{1}{N_r}\sum_{n_r=1}^{N_r}\sum_{n'_r=1}^{N_r}e^{\frac{j2\pi \hspace{-0.04cm}\left[l(n_r\hspace{-0.04cm}-\hspace{-0.04cm}1)\hspace{-0.04cm}-\hspace{-0.04cm}l'(n'_r\hspace{-0.04cm}-\hspace{-0.04cm}1)\right]}{N_r}}\nonumber\\
&\sum_{n_t=1}^{N_t}\hspace{-0.08cm}\int_{y_{n_t}-\frac{L}{2}}^{y_{n_t}+\frac{L}{2}}\hspace{-0.15cm}\int_{y_{n_t}-\frac{L}{2}}^{y_{n_t}+\frac{L}{2}}\hspace{-0.1cm}
\begin{bmatrix}G^{xy}(\boldsymbol{\mathrm r}_{n_r},\boldsymbol{\mathrm s}) \\ G^{yy}(\boldsymbol{\mathrm r}_{n_r},\boldsymbol{\mathrm s}) \\ G^{zy}(\boldsymbol{\mathrm r}_{n_r},\boldsymbol{\mathrm s})\end{bmatrix}\hspace{-0.15cm}\begin{bmatrix}G^{xy}(\boldsymbol{\mathrm r}_{n'_r},\boldsymbol{\mathrm s}') \\ G^{yy}(\boldsymbol{\mathrm r}_{n'_r},\boldsymbol{\mathrm s}') \\ G^{zy}(\boldsymbol{\mathrm r}_{n'_r},\boldsymbol{\mathrm s}')\end{bmatrix}^H\hspace{-0.35cm}{\mathrm d}s^y{\mathrm d}s'^y.
\label{eq:R_y2}
\end{align}
\end{small}
\hspace{-0.18cm}
Eq.~\eqref{eq:E_autocorr2} can also be rewritten to a discrete form by substituting Eq.~\eqref{eq:E_autocorr_D} into \eqref{eq:R_y} as follows:
\begin{small}
\begin{align}
{\boldsymbol{\mathrm R}}_{\boldsymbol{\mathrm y}}(l,l')\hspace{-0.1cm}&=\hspace{-0.1cm}\frac{1}{N_rN^2_l}\sum_{n_r=1}^{N_r}\sum_{n'_r=1}^{N_r}e^{\frac{j2\pi \hspace{-0.04cm}\left[l(n_r\hspace{-0.04cm}-\hspace{-0.04cm}1)\hspace{-0.04cm}-\hspace{-0.04cm}l'(n'_r\hspace{-0.04cm}-\hspace{-0.04cm}1)\right]}{N_r}}\nonumber\\
&\quad\sum_{n_t=1}^{N_t}\sum_{n_l = 1}^{N_l}\sum_{n'_l = 1}^{N_l}\begin{bmatrix}G^{xy}(\boldsymbol{\mathrm r}_{n_r},\boldsymbol{\mathrm s}_{n_t,n_l}) \\ G^{yy}(\boldsymbol{\mathrm r}_{n_r},\boldsymbol{\mathrm s}_{n_t,n_l}) \\ G^{zy}(\boldsymbol{\mathrm r}_{n_r},\boldsymbol{\mathrm s}_{n_t,n_l})\end{bmatrix}\hspace{-0.15cm}\begin{bmatrix}G^{xy}(\boldsymbol{\mathrm r}_{n'_r},\boldsymbol{\mathrm s}_{n_t,n'_l}) \\ G^{yy}(\boldsymbol{\mathrm r}_{n'_r},\boldsymbol{\mathrm s}_{n_t,n'_l}) \\ G^{zy}(\boldsymbol{\mathrm r}_{n'_r},\boldsymbol{\mathrm s}_{n_t,n'_l})\end{bmatrix}^H.
\label{eq:R_y_D}
\end{align}
\end{small}

\begin{figure}[!h]
\centering
%\vspace{-15pt}
\subfigure[$x$-$x$ components.]{
\begin{minipage}{0.27\linewidth}
\centering
\includegraphics[scale=0.19]{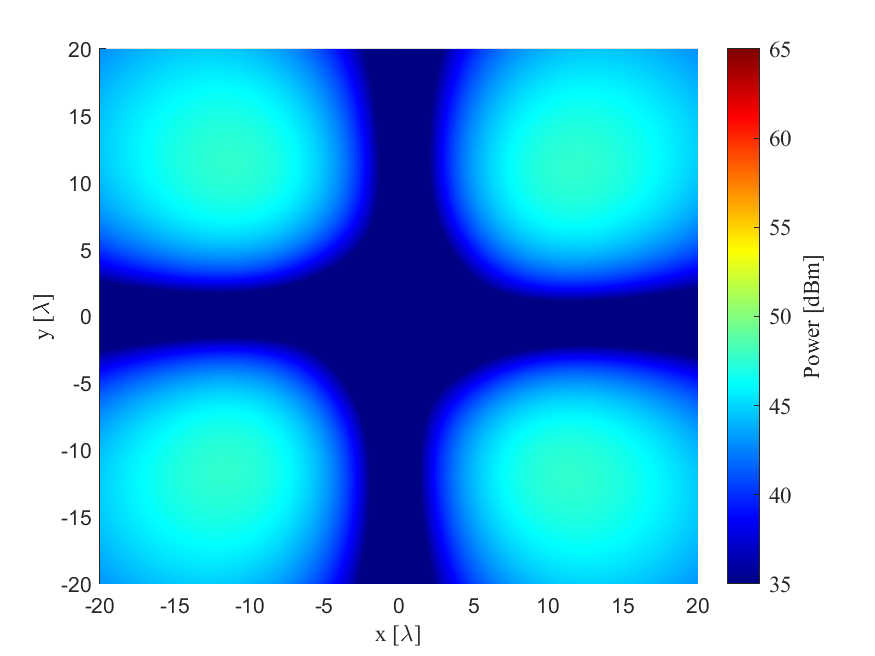}
%\vspace{-30pt}
\end{minipage}
}
\subfigure[$x$-$y$ components.]{
\begin{minipage}{0.27\linewidth}
\centering
\includegraphics[scale=0.19]{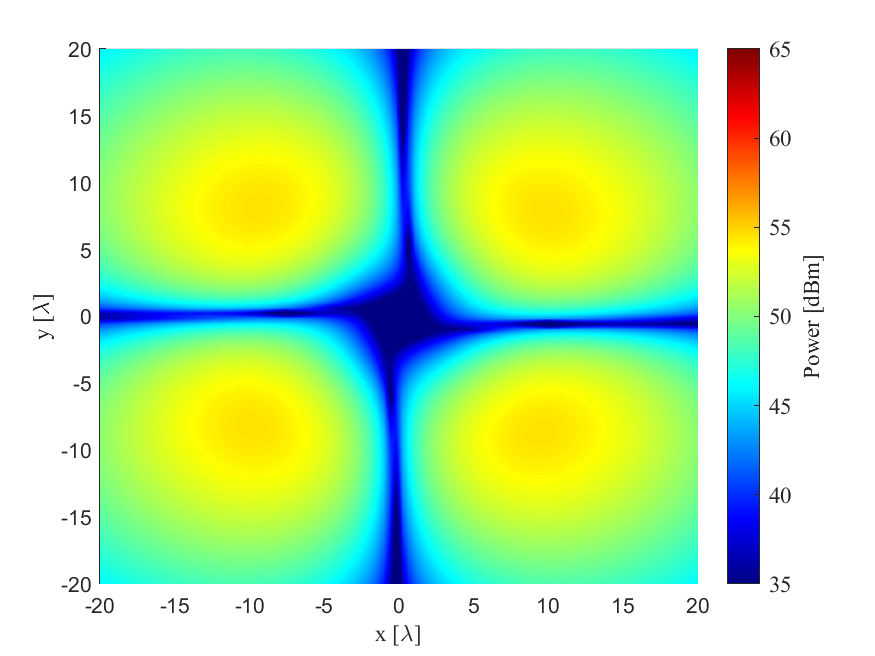}
%\vspace{-30pt}
\end{minipage}
}
\subfigure[$x$-$z$ components.]{
\begin{minipage}{0.27\linewidth}
\centering
\includegraphics[scale=0.19]{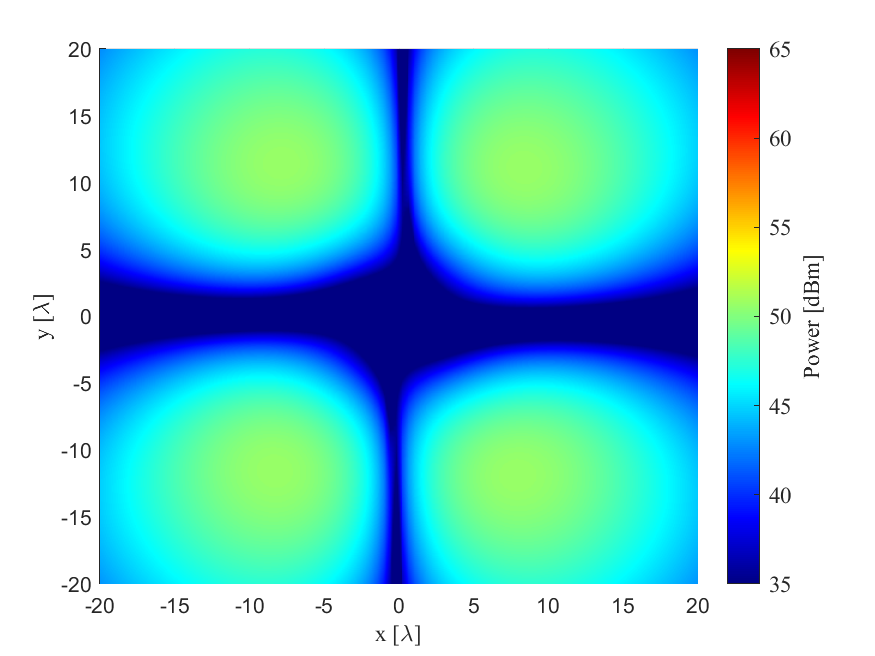}
%\vspace{-30pt}
\end{minipage}
}
\subfigure[$y$-$x$ components.]{
\begin{minipage}{0.27\linewidth}
\centering
\includegraphics[scale=0.19]{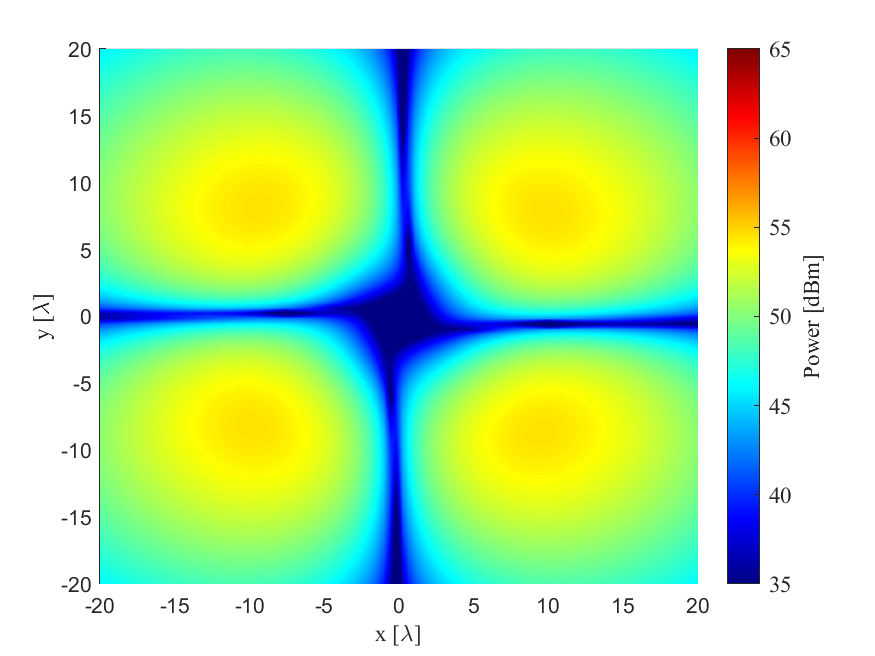}
%\vspace{-30pt}
\end{minipage}
}
\subfigure[$y$-$y$ components.]{
\begin{minipage}{0.27\linewidth}
\centering
\includegraphics[scale=0.19]{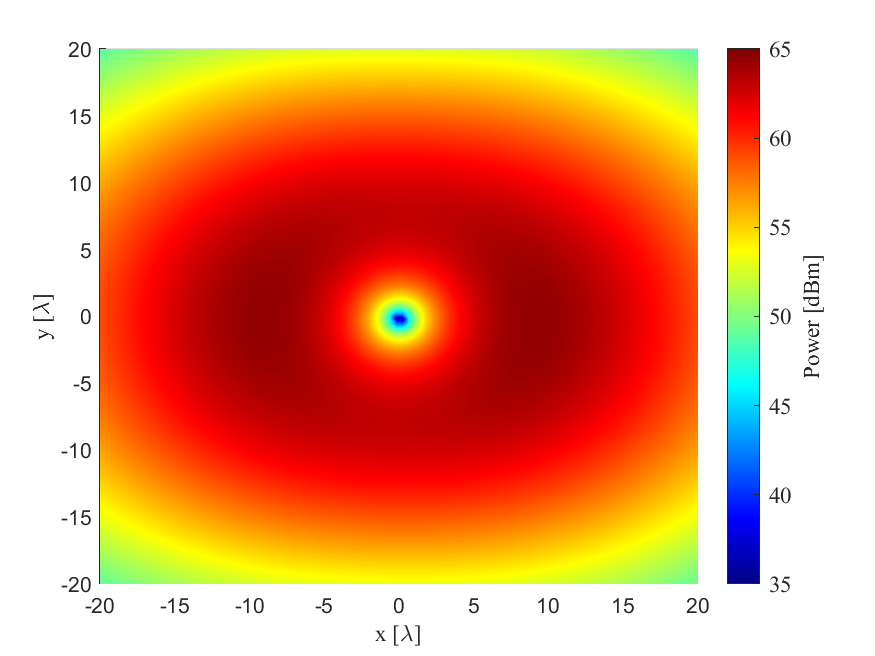}
%\vspace{-30pt}
\end{minipage}
}
\subfigure[$y$-$z$ components.]{
\begin{minipage}{0.27\linewidth}
\centering
\includegraphics[scale=0.19]{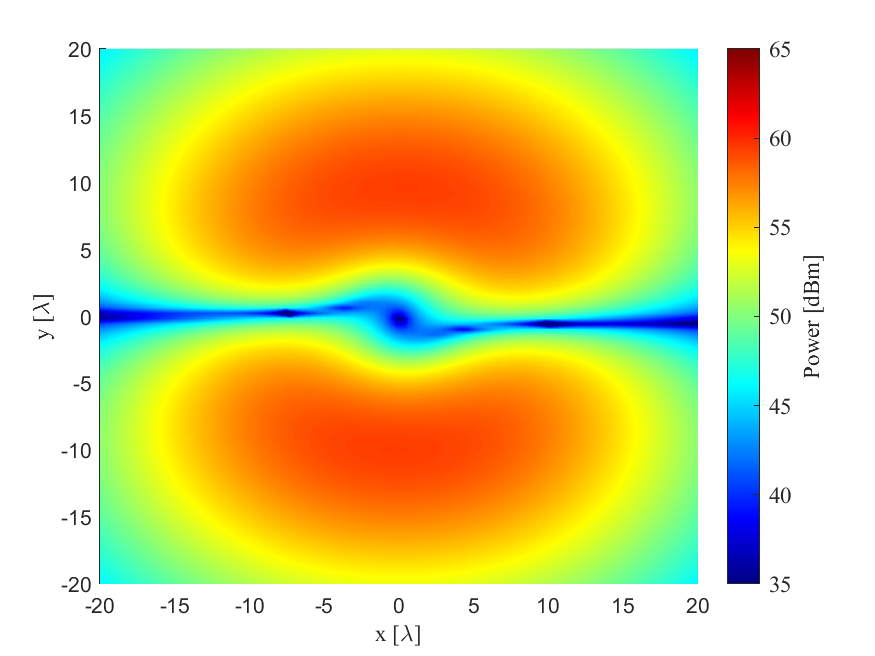}
%\vspace{-30pt}
\end{minipage}
}
\subfigure[$z$-$x$ components.]{
\begin{minipage}{0.27\linewidth}
\centering
\includegraphics[scale=0.19]{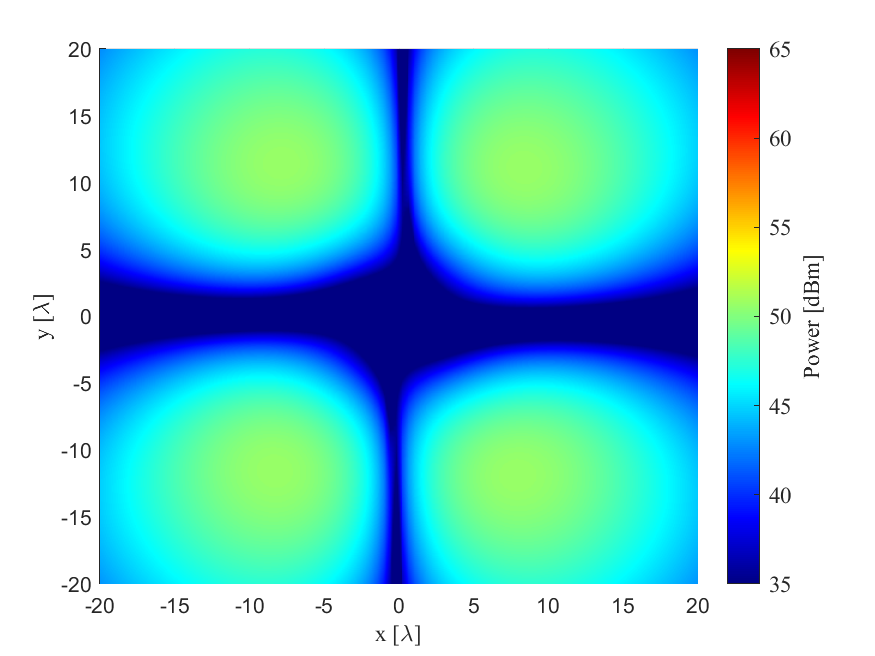}
%\vspace{-30pt}
\end{minipage}
}
\subfigure[$z$-$y$ components.]{
\begin{minipage}{0.27\linewidth}
\centering
\includegraphics[scale=0.19]{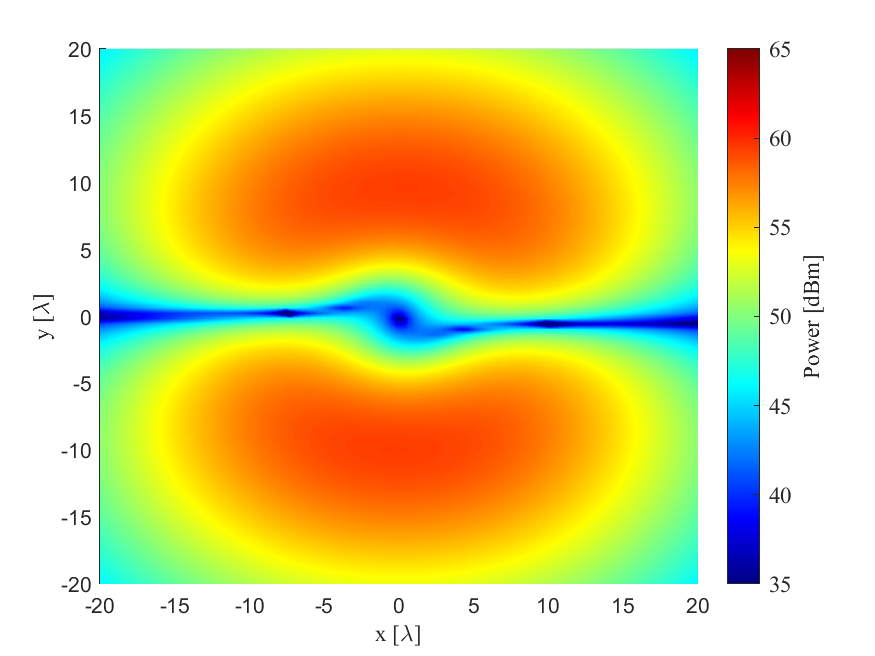}
%\vspace{-30pt}
\end{minipage}
}
\subfigure[$z$-$z$ components.]{
\begin{minipage}{0.27\linewidth}
\centering
\includegraphics[scale=0.19]{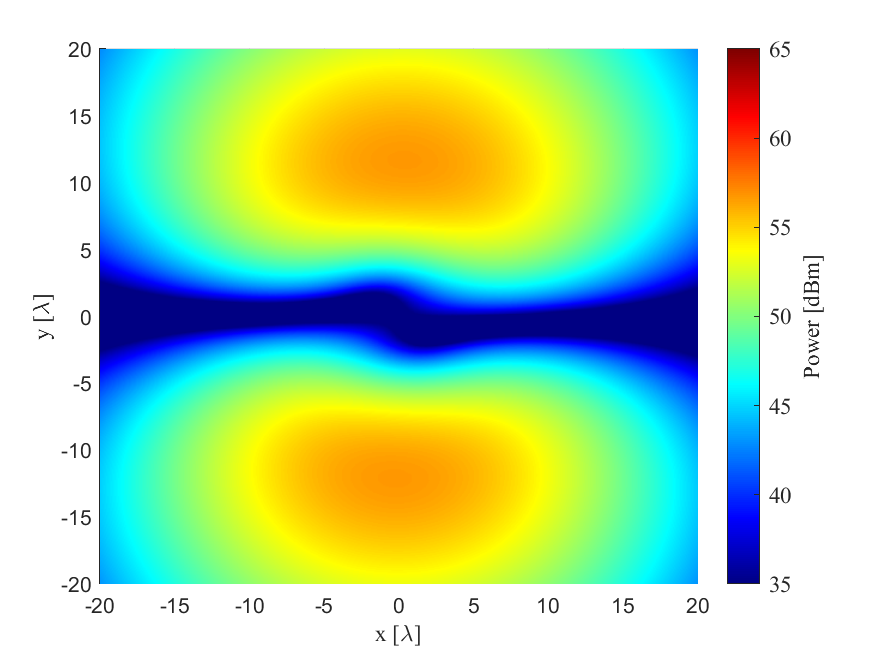}
%\vspace{-30pt}
\end{minipage}
}
\centering
%\vspace{-10pt}
\caption{Powers for autocorrelation functions of OAM-mode $1$.} \label{fig:autocorrelationMode1}
%\vspace{-15pt}
\end{figure}
We also plot the power of the autocorrelation functions for OAM mode $1$ in Fig.~\ref{fig:autocorrelationMode1}, using the same settings as in Fig.~\ref{fig:TxObserver} based on Eq.~\eqref{eq:R_y_D}. As shown, the power for the autocorrelation of the $y$-$y$ components is significantly higher than that for the autocorrelations of other component combinations.

\subsection{Channel Capacity Derivation}
Since the current densities of the sources are aligned along the $y$-axis, the autocorrelation in the $y$ direction is significantly greater than in any other direction. Thus, in the following sections, we use the cross-correlation between the received signals of different OAM modes and the autocorrelation of the same OAM mode along the $y$-axis to derive the mutual information and channel capacity. We denote by $R_{\boldsymbol{\mathrm y}}(l,l')$ the $y$ component cross-correlation function between the received signals of the $l$th and $l'$th OAM modes. $R_{\boldsymbol{\mathrm y}}(l,l')$ can be given as follows:
\begin{align}
R_{\boldsymbol{\mathrm y}}(l,l')\hspace{-0.1cm}&=\hspace{-0.1cm}\frac{1}{N_r}\sum_{n_r=1}^{N_r}\sum_{n'_r=1}^{N_r}\sum_{n_t=1}^{N_t}e^{\frac{j2\pi \hspace{-0.04cm}\left[l(n_r\hspace{-0.04cm}-\hspace{-0.04cm}1)\hspace{-0.04cm}-\hspace{-0.04cm}l'(n'_r\hspace{-0.04cm}-\hspace{-0.04cm}1)\right]}{N_r}}\nonumber\\
&\sum_{n_t=1}^{N_t}\hspace{-0.08cm}\int_{y_{n_t}-\frac{L}{2}}^{y_{n_t}+\frac{L}{2}}\hspace{-0.15cm}\int_{y_{n_t}-\frac{L}{2}}^{y_{n_t}+\frac{L}{2}}\hspace{-0.1cm}
G^{yy}(\boldsymbol{\mathrm r}_{n_r},\boldsymbol{\mathrm s})\hspace{-0.15cm}\left[G^{yy}(\boldsymbol{\mathrm r}_{n'_r},\boldsymbol{\mathrm s}')\right]^*\hspace{-0.2cm}{\mathrm d}s^y{\mathrm d}s'^y.
\label{eq:R_y3}
\end{align}
And the discrete form is given as follows:
\begin{align}
R_{\boldsymbol{\mathrm y}}(l,l')\hspace{-0.1cm}&=\hspace{-0.1cm}\frac{1}{N_rN^2_l}\sum_{n_r=1}^{N_r}\sum_{n'_r=1}^{N_r}\sum_{n_t=1}^{N_t}e^{\frac{j2\pi \hspace{-0.04cm}\left[l(n_r\hspace{-0.04cm}-\hspace{-0.04cm}1)\hspace{-0.04cm}-\hspace{-0.04cm}l'(n'_r\hspace{-0.04cm}-\hspace{-0.04cm}1)\right]}{N_r}}\nonumber\\
&\quad\sum_{n_t=1}^{N_t}\sum_{n_l = 1}^{N_l}\sum_{n'_l = 1}^{N_l}G^{yy}(\boldsymbol{\mathrm r}_{n_r},\boldsymbol{\mathrm s}_{n_t,n_l})\left[G^{yy}(\boldsymbol{\mathrm r}_{n'_r},\boldsymbol{\mathrm s}_{n_t,n'_l})\right]^*.
\label{eq:R_y3_D}
\end{align}
Therefore, the OAM channel correlation matrix, denoted by ${\boldsymbol{\mathrm R}}_{\rm OAM}$, can be given as follows:
%\begin{align}
%{\boldsymbol{\mathrm R}}_{\rm OAM} =
%\begin{bmatrix}
%R_{\boldsymbol{\mathrm y}}(0,0) &\cdots &R_{\boldsymbol{\mathrm y}}(0,N_t-1)\\
%\vdots &\ddots &\vdots\\
%R_{\boldsymbol{\mathrm y}}(N_r-1,0) &\cdots &R_{\boldsymbol{\mathrm y}}(N_r-1,N_t-1)
%\end{bmatrix}.
%\label{eq:R_OAM}
%\end{align}
\begin{align}
{\boldsymbol{\mathrm R}}_{\rm OAM} =
\begin{bmatrix}
R_{\boldsymbol{\mathrm y}}(0,0) &\cdots &R_{\boldsymbol{\mathrm y}}(0,N_r-1)\\
\vdots &\ddots &\vdots\\
R_{\boldsymbol{\mathrm y}}(N_r-1,0) &\cdots &R_{\boldsymbol{\mathrm y}}(N_r-1,N_r-1)
\end{bmatrix}.
\label{eq:R_OAM}
\end{align}
Thus, when the transmitted power is evenly distributed to each mode, the channel capacity of the OAM wireless communication with additive white gaussian noise (AWGN), denoted by $C$, can be given as follows:
\begin{align}
%C={\rm log}_2\left|\boldsymbol{\mathrm I}_{N_r}+\frac{P/N_t}{N_0}{\boldsymbol{\mathrm R}}_{\rm OAM}{\boldsymbol{\mathrm R}}^H_{\rm OAM} \right|,
%C={\rm log}_2\left|\boldsymbol{\mathrm I}_{N_r}+\frac{P/N_t}{N_0}\left[{\boldsymbol{\mathrm R}}_{\rm OAM}, \boldsymbol{0}_{N_r\times(N_r-N_t)}\right]\right|,
C={\rm log}_2\left|\boldsymbol{\mathrm I}_{N_r}+\frac{P/N_t}{N_0}{\boldsymbol{\mathrm R}}_{\rm OAM}\right|,
\label{eq:C}
\end{align}
where $P$ denotes the transmitted power and $N_0$ represents the power of the AWGN.

\section{Numerical Results}\label{sec:numericalResults}
In this section, we give numerical results to validate the channel capacity enhancement via exploration under EIT framework. We also analyze the impact of various parameters on system performance, including including numbers of transmitting and receiving mode, transmit and receive UCA distance, and UCA radii.

\begin{figure}[htbp]
\centering
%\vspace{-10pt}
\includegraphics[scale=0.55]{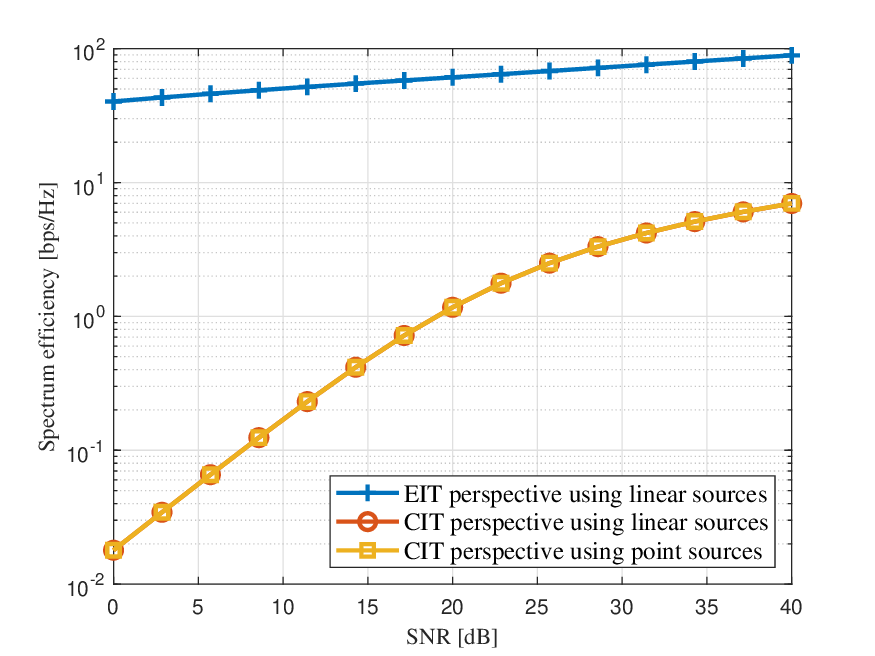}
%\vspace{-10pt}
\caption{Enhancement of channel capacity for OAM-based wireless communication from an EIT perspective.} \label{fig:EitOamEnhance}
%\vspace{-5pt}
\end{figure}
Figure~\ref{fig:EitOamEnhance} in the next page illustrates the channel capacities of OAM-based wireless communications calculated from both EIT and CIT perspectives, based on the system model presented in Section~\ref{sec:systemModel}. Additionally, we plot the channel capacity of OAM system with transmit and receive UCAs composed of point sources for comparison. The number of transmitting and receiving modes is set to $8$, and the operating frequency is $5.8$ GHz, with a fixed distance of $200\lambda$ ($10344.8$ mm) between the transmit and receive UCAs. Both UCAs are equipped with $8$ sources, where each linear source in the transmit UCA consists of $10$ small feeds. The current density of each feed in the transmit linear source is normalized by dividing it by $\sqrt{80}$, and the length of each linear source is $0.5\lambda$ ($25.9$ mm). For the point sources, each consists of one small feed, and the current density is normalized by dividing it by $\sqrt{8}$. The radii of both the transmit and receive UCAs are set to $2\lambda/\pi$ ($32.9$ mm). As depicted in Fig.~\ref{fig:EitOamEnhance}, the EIT method outperforms CIT methods (both with linear and point sources) in terms of spectrum efficiency, particularly at lower SNR values. Among the CIT methods, using linear sources or point sources does not change the channel capacity.

\begin{figure}[htbp]
\centering
%\vspace{-10pt}
\includegraphics[scale=0.55]{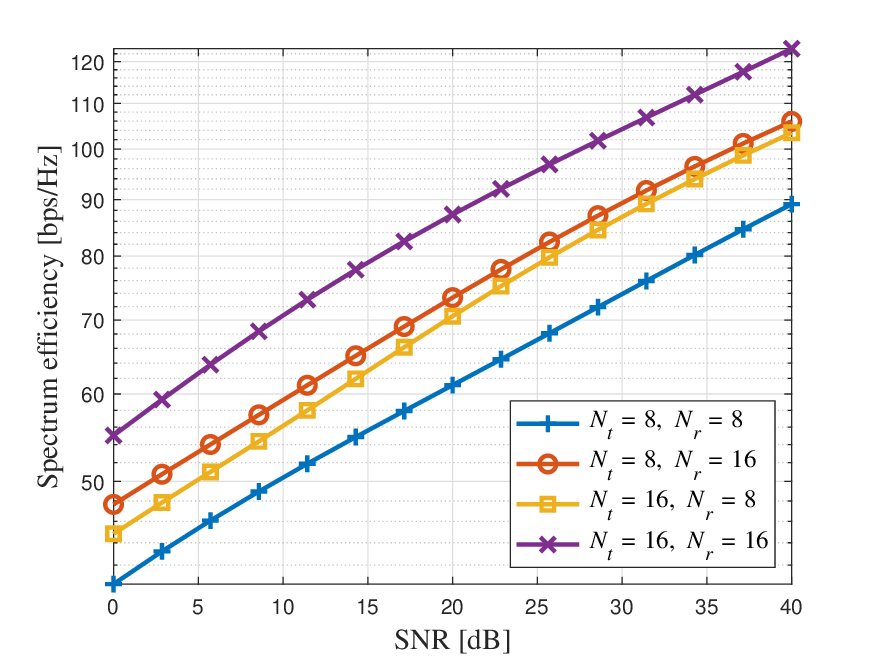}
%\vspace{-10pt}
\caption{OAM channel capacities with different numbers of transmitting and receiving mode.} \label{fig:capacityNrNt}
%\vspace{-5pt}
\end{figure}
Figure~\ref{fig:capacityNrNt} illustrates the channel capacities of the OAM-based wireless communication system, where the number of transmitting and receiving modes alternates between $8$ and $16$. The operating frequency is also set at $5.8$ GHz, and the distance between the transmit and receive UCAs is also fixed at $200\lambda$ ($10344.8$ mm). The transmit and receive UCAs are equipped with either $8$ or $16$ antennas. The transmit UCA consists of $10$ small feeds. The current density for each feed of the transmit UCA is normalized by dividing it by $\sqrt{10N_t}$. The length of each transmit source is set to $0.5\lambda$ ($25.9$ mm), while the radii of the transmit and receive UCAs are given by $N_t\lambda/(4\pi)$ and $N_r\lambda/(4\pi)$, respectively. Figure~\ref{fig:capacityNrNt} compares the spectrum efficiency under four different configurations: $(N_t = 8, N_r = 8)$, $(N_t = 8, N_r = 16)$, $(N_t = 16, N_r = 8)$, and $(N_t = 16, N_r = 16)$. The system's spectrum efficiency increases with SNR across all configurations. As shown in Fig.~\ref{fig:capacityNrNt}, the configuration with $N_t = 16$ and $N_r = 16$ achieves the highest spectrum efficiency, surpassing $120$ bps/Hz at SNR$=40$ dB. The performances of the $(N_t = 8, N_r = 16)$ and $(N_t = 16, N_r = 8)$ configurations are nearly identical, though both are lower than that of $(N_t = 16, N_r = 16)$, indicating that increasing neither transmitting and receiving OAM modes can enhance capacity. Notably, increasing the number of receiving OAM modes offers a greater improvement in capacity compared to increasing the number of transmitting OAM modes. The $(N_t = 8, N_r = 8)$ configuration shows the lowest spectrum efficiency, attributed to the fewer OAM modes used for both transmission and reception.

%\begin{figure}[htbp]
%\centering
%%\vspace{-10pt}
%\includegraphics[scale=0.55]{pics//capacityNlNrNt.eps}
%%\vspace{-10pt}
%\caption{EIT-based OAM channel capacities with different small feed number for each source.} \label{fig:capacityNlNrNt}
%%\vspace{-5pt}
%\end{figure}
%In Fig.~\ref{fig:capacityNlNrNt}, the number of small feeds for both the transmit and receive UCAs is varied from $1$ to $30$, while all other parameters remain the same as in Fig.~\ref{fig:capacityNrNt}. The results in Fig.~\ref{fig:capacityNlNrNt} indicate that varying the number of small feeds has a negligible impact on the channel capacity, suggesting that the feed number does not significantly influence the overall system performance.

\begin{figure}[htbp]
\centering
%\vspace{-10pt}
\includegraphics[scale=0.55]{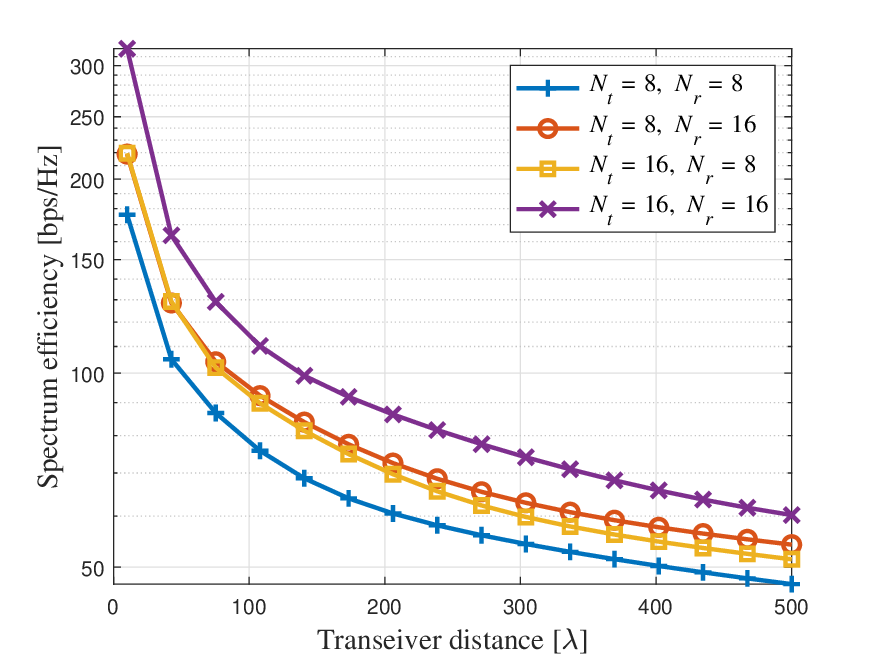}
%\vspace{-10pt}
\caption{OAM channel capacities with different transmit and receive UCA distances.} \label{fig:capacityDistanceNrNt}
%\vspace{-5pt}
\end{figure}
Figure~\ref{fig:capacityDistanceNrNt} illustrates the OAM channel capacities with the transmit and receive UCA distance varying from $10$ to $500\lambda$ ($517.2$ to $25862$ mm) for four different configurations of transmitting and receiving modes: $(N_t = 8, N_r = 8)$, $(N_t = 8, N_r = 16)$, $(N_t = 16, N_r = 8)$, and $(N_t = 16, N_r = 16)$. We set SNR $= 20$ dB. All other parameters remain the same as in Fig.~\ref{fig:capacityNrNt}. As shown in Fig.~\ref{fig:capacityDistanceNrNt}, all configurations experience a reduction in spectrum efficiency as the transceiver distance increases, which is expected due to propagation losses over longer distances. Among the configurations, $(N_t = 16, N_r = 16)$ consistently achieves the highest spectrum efficiency across all distances. Conversely, the $(N_t = 8, N_r = 8)$ configuration has the lowest spectrum efficiency. The performances of the $(N_t = 8, N_r = 16)$ and $(N_t = 16, N_r = 8)$ configurations remain very similar and closely match that of $(N_t = 16, N_r = 16)$. %The results shown in Fig.~\ref{fig:capacityDistanceNrNt} are consistent with those in Fig.~\ref{fig:capacityNrNt}.

\begin{figure}[htbp]
\centering
%\vspace{-15pt}
\subfigure[Transmit UCA radius impact.]{
\begin{minipage}{1\linewidth}
\centering
\includegraphics[scale=0.55]{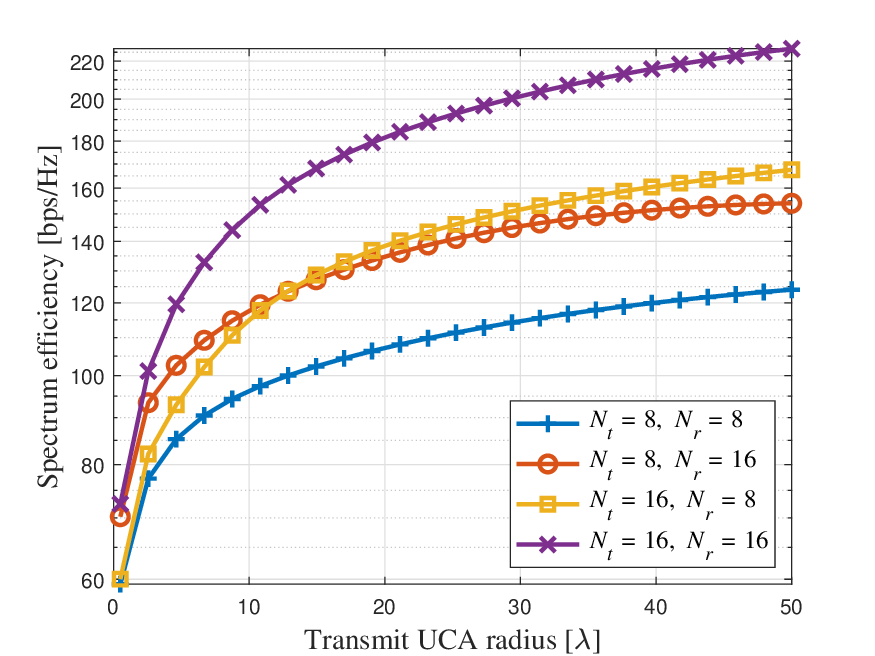}\label{fig:capacityRtNrNt}
%\vspace{-30pt}
\end{minipage}
}\\
\subfigure[Receive UCA radius impact.]{
\begin{minipage}{1\linewidth}
\centering
\includegraphics[scale=0.55]{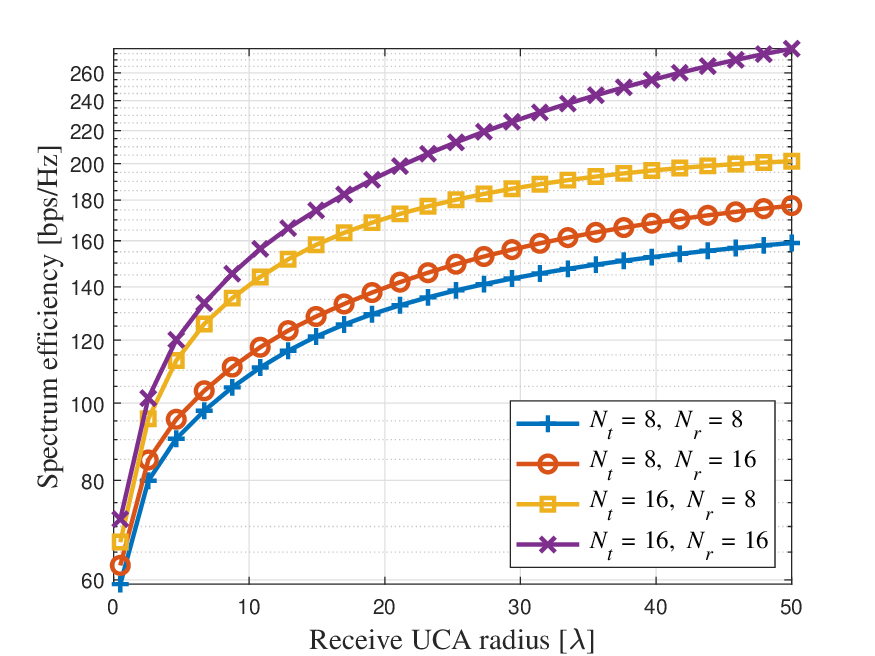}\label{fig:capacityRrNrNt}
%\vspace{-30pt}
\end{minipage}
}
\centering
%\vspace{-10pt}
\caption{OAM channel capacities with different transmit and receive UCA radii.} \label{fig:capacityRtRrNrNt}
%\vspace{-10pt}
\end{figure}
Figure~\ref{fig:capacityRtRrNrNt} illustrates the impact of different transmit and receive UCA radii on the channel capacity of the OAM-based wireless communication system. The results are also shown for four configurations: $(N_t = 8, N_r = 8)$, $(N_t = 8, N_r = 16)$, $(N_t = 16, N_r = 8)$, and $(N_t = 16, N_r = 16)$. Subplots \ref{fig:capacityRtNrNt} and \ref{fig:capacityRrNrNt} depict the effect of varying the transmit UCA radius and the receive UCA radius from $1$ to $50\lambda$ ($51.7$ to $2586.2$ mm), respectively, on the system's spectral efficiency. The SNR is set to $20$ dB. All other parameters remain the same as in Fig.~\ref{fig:capacityNrNt}. As shown in Fig.~\ref{fig:capacityRtNrNt}, all configurations exhibit an increase in spectrum efficiency as the transmit UCA radius increases. Among the configurations, $(N_t = 16, N_r = 16)$ consistently achieves the highest spectrum efficiency across all transmit UCA radii. In contrast, the $(N_t = 8, N_r = 8)$ configuration demonstrates the lowest spectrum efficiency. The performances of $(N_t = 8, N_r = 16)$ and $(N_t = 16, N_r = 8)$ are comparable, although both remain lower than $(N_t = 16, N_r = 16)$ as the UCA radius increases. Notably, $(N_t = 16, N_r = 8)$ exhibits lower spectrum efficiency than $(N_t = 8, N_r = 16)$ when the radius is below $15\lambda$, but surpasses it at larger radii.

Figure~\ref{fig:capacityRrNrNt} shows that all configurations exhibit a similar trend of increasing spectrum efficiency as the receive UCA radius increases. The $(N_t = 16, N_r = 16)$ configuration achieves the highest spectrum efficiency, followed by $(N_t = 16, N_r = 8)$. While the $(N_t = 8, N_r = 16)$ configuration has a lower spectrum efficiency than $(N_t = 16, N_r = 8)$, it grows more rapidly as the receive UCA radius increases. The $(N_t = 8, N_r = 8)$ configuration remains the least efficient across all UCA radii.

%\begin{figure}[htbp]
%\centering
%\vspace{-10pt}
%\subfigure[OAM-mode 1 Power.]{
%\begin{minipage}{0.45\linewidth}
%\centering
%\includegraphics[scale=0.3105]{pics//talbotPhaseYmode1.eps}\label{fig:mode1Power}
%%\vspace{-20pt}
%\end{minipage}
%}
%\subfigure[OAM-mode 1 Phase.]{
%\begin{minipage}{0.45\linewidth}
%\centering
%\includegraphics[scale=0.3105]{pics//talbotPowerYmode1.eps}\label{fig:mode1Phase}
%%\vspace{-20pt}
%\end{minipage}
%}
%\centering
%%\vspace{-5pt}
%\caption{Talbot effect generated by UCAs with $6$ elements.}\label{fig:TalbotEffect_OAM}
%%\vspace{-10pt}
%\end{figure}

\section{Simulations}\label{sec:simulations}
In this section, we provide simulations in ANSYS high frequency structure simulator (HFSS). We first give our proposed OAM-NFC simulation model in HFSS. Then, we simulate the electric fields, complex magnitudes, and phase to validate the derivations of electric field in Section.~\ref{sec:OAM_EIT}.

\begin{figure}[htbp]
\centering
%\vspace{-10pt}
\includegraphics[scale=0.43]{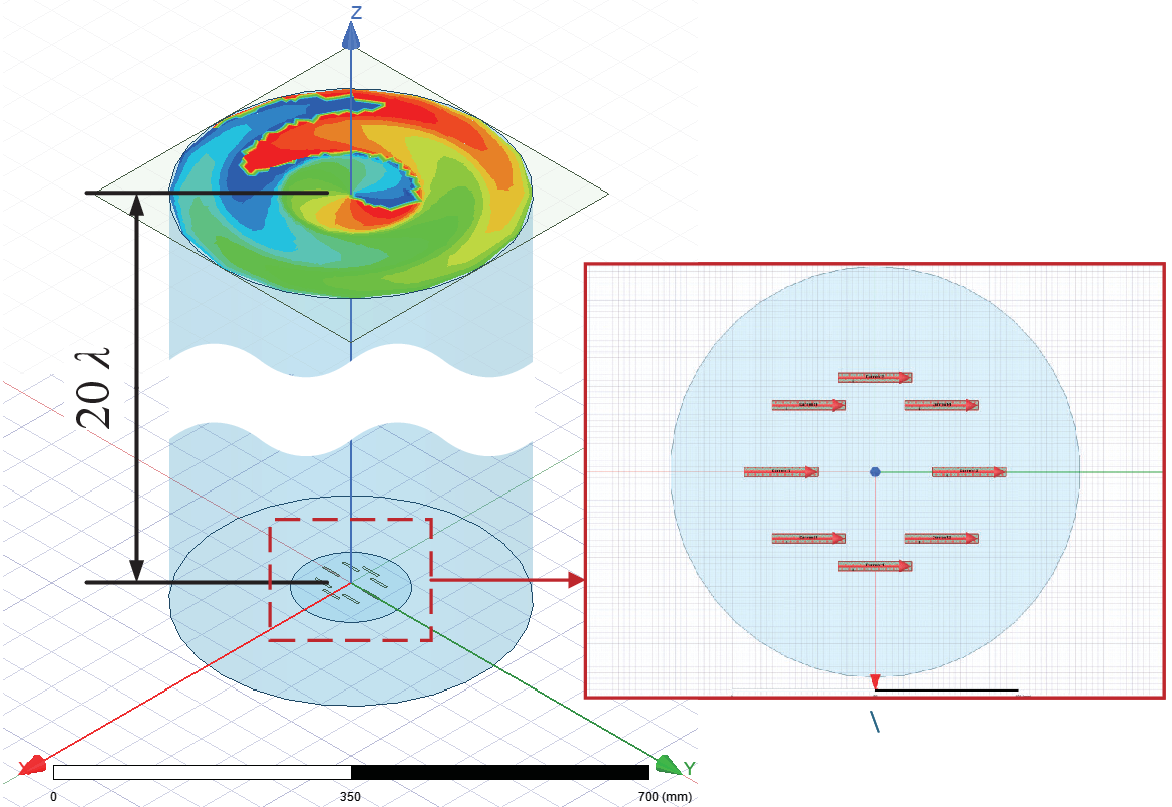}
%\vspace{-10pt}
\caption{UCA model and excitations in HFSS.} \label{fig:HfssUCA}
%\vspace{-5pt}
\end{figure}
Figure~\ref{fig:HfssUCA} illustrates the UCA model and excitations simulated in HFSS, where the system operates at a frequency of $5.8$ GHz. The distance between the transmit UCA and the observer plane is fixed at $20\lambda$. The radius of the transmit UCA is set to $2\lambda/\pi$, which equals to approximately $32.9$ mm. The UCA is configured with $8$ linear sources, all aligned along the $y$-axis. Each linear source has a length of $\lambda/2$ ($25.9$ mm) and a width of $\lambda/16$ ($3.2$ mm). The currents are also aligned along the $y$-axis and are normalized by set the magnitudes as $1/\sqrt{8}$ A.

\begin{figure}[htbp]
\centering
%\vspace{-15pt}
\subfigure[OAM-mode $-1$.]{
\begin{minipage}{0.45\linewidth}
\centering
\includegraphics[scale=0.2]{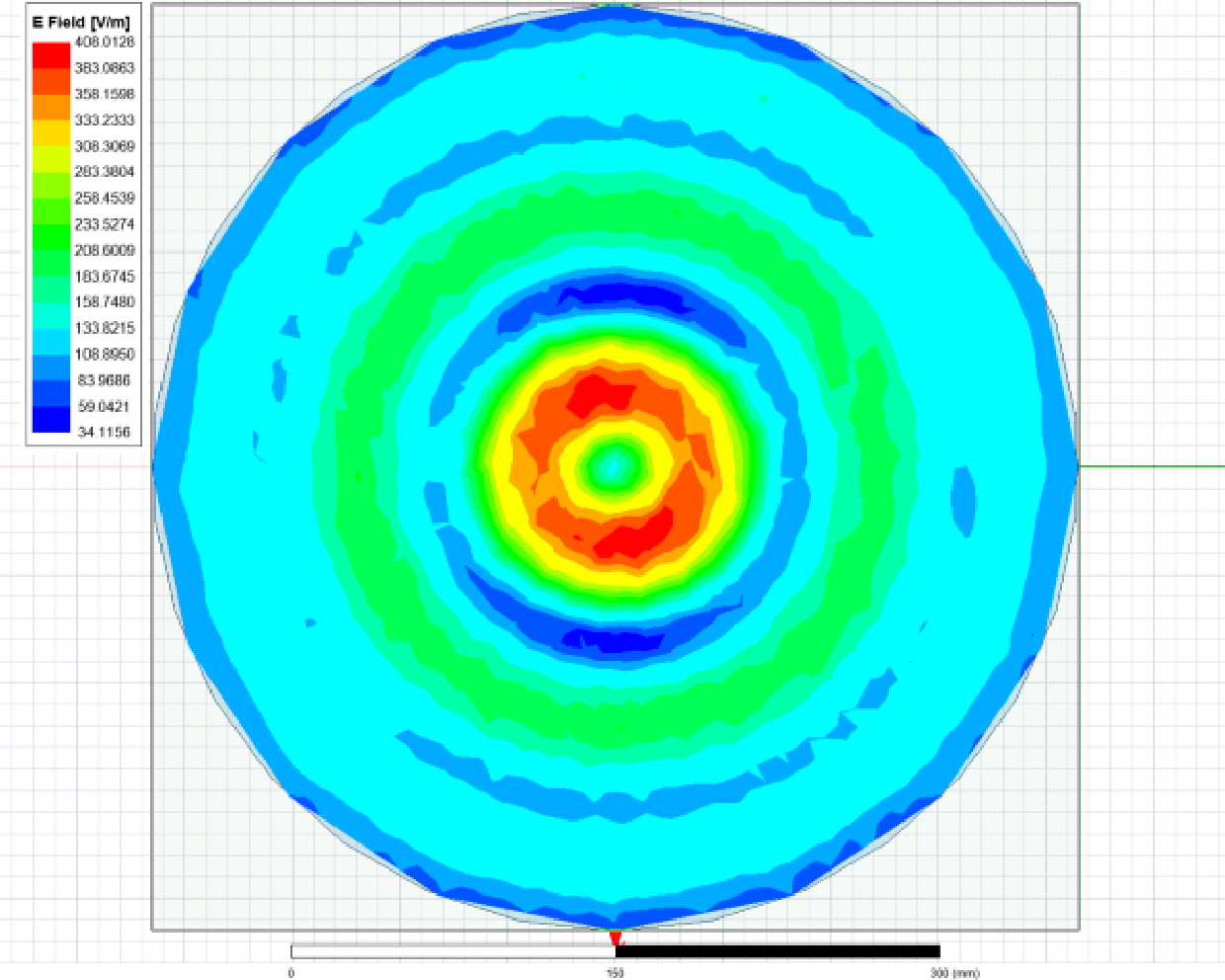}
%\vspace{-30pt}
\end{minipage}
}
\subfigure[OAM-mode $0$.]{
\begin{minipage}{0.45\linewidth}
\centering
\includegraphics[scale=0.2]{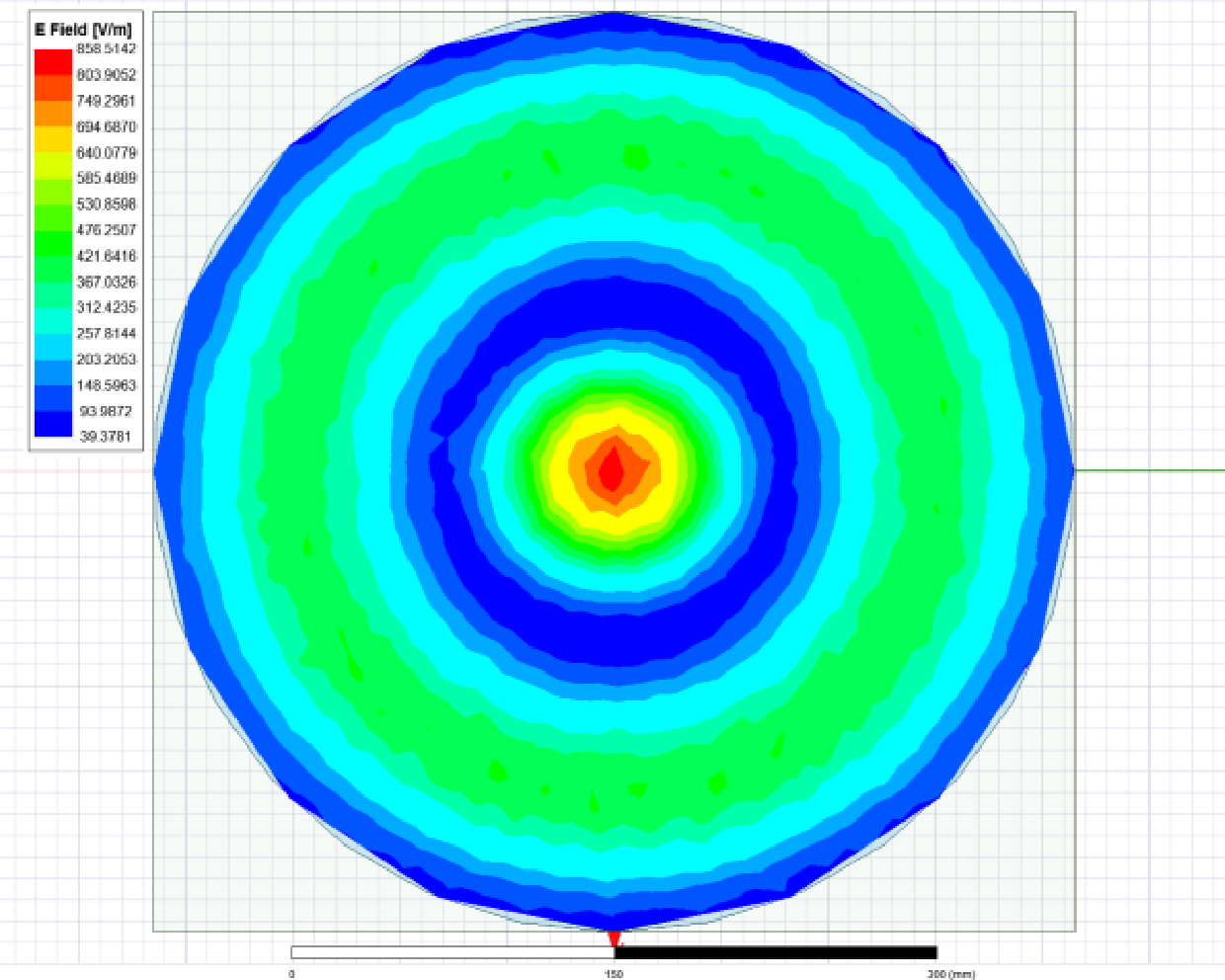}
%\vspace{-30pt}
\end{minipage}
}
\subfigure[OAM-mode $1$.]{
\begin{minipage}{0.45\linewidth}
\centering
\includegraphics[scale=0.2]{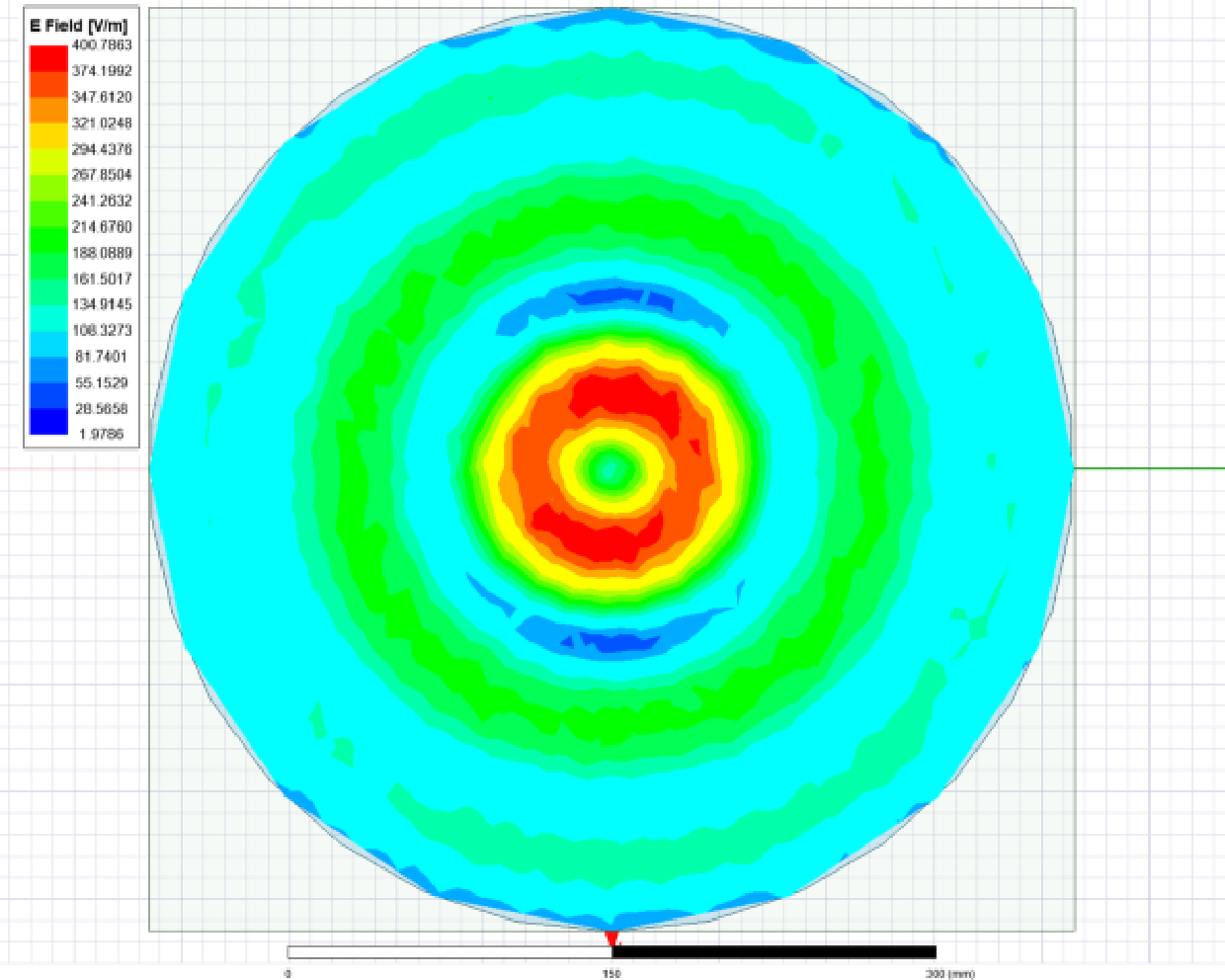}
%\vspace{-30pt}
\end{minipage}
}
\subfigure[OAM-mode $2$.]{
\begin{minipage}{0.45\linewidth}
\centering
\includegraphics[scale=0.2]{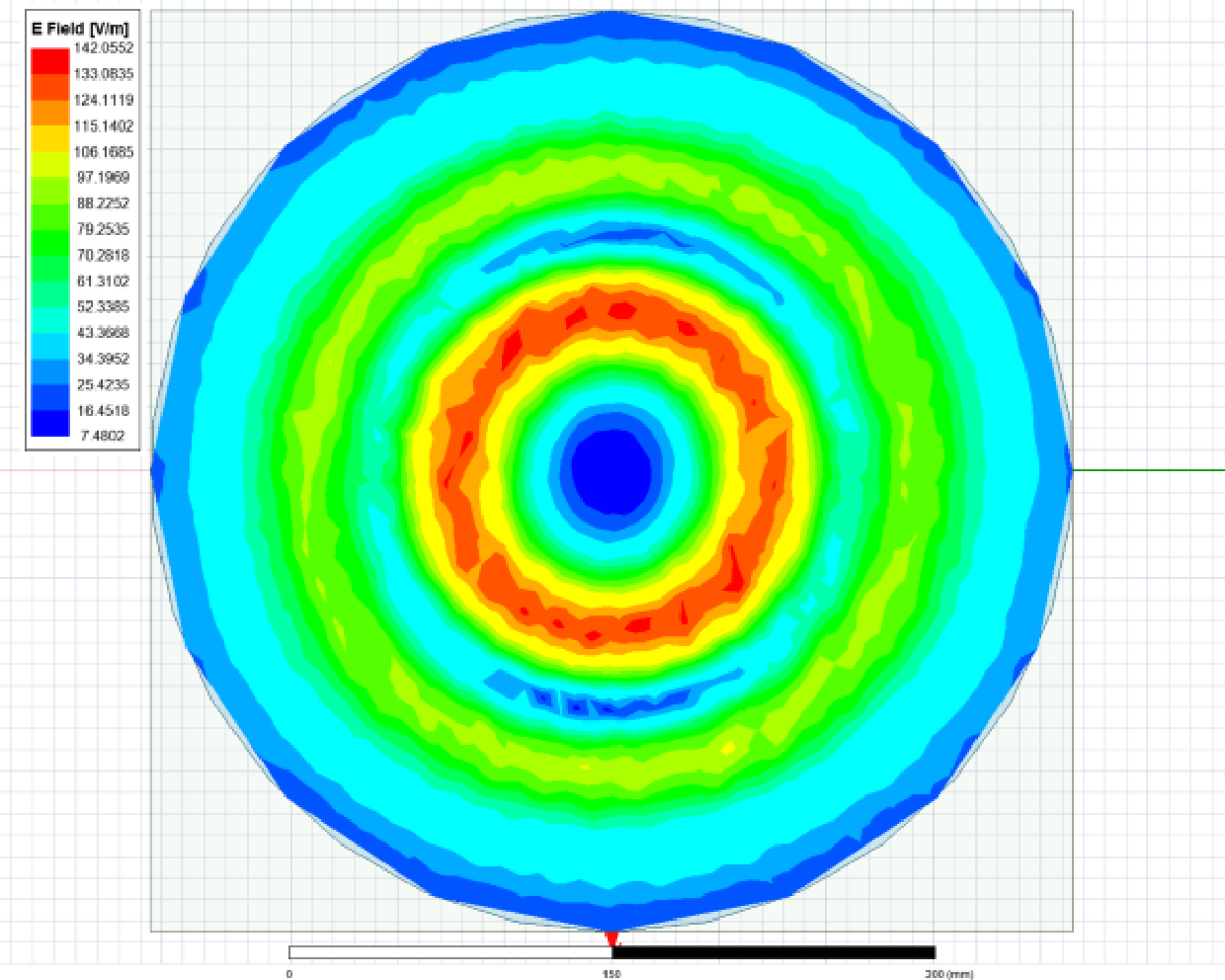}
%\vspace{-30pt}
\end{minipage}
}
\centering
%\vspace{-10pt}
\caption{Complex magnitudes of OAM-mode $-1$, $0$, $1$, and $2$ simulated by HFSS.} \label{fig:eFiled_HFSS}
%\vspace{-15pt}
\end{figure}
Figure~\ref{fig:eFiled_HFSS} in the next page illustrates the complex magnitudes of OAM modes $-1$, $0$, $1$, and $2$ simulated by HFSS. The width of the observer plane is set to $430$ mm. All four subplots represent the field distributions in a circular symmetry, with colors changing from red (representing areas of higher field magnitude) to blue (indicating lower magnitudes). The general pattern reveals concentric rings in each plot, highlighting the hollow divergence of OAM beams. The most evident difference lies in the central region. For OAM-mode $0$, the field intensity peaks at the center, forming a Gaussian-like distribution, while the other non-zero modes exhibit a central null. OAM modes $-1$ and $1$ are quite similar, each showing a high-power ring around the central null, though the phase and intensity distribution are subtly different due to their opposite OAM modes. OAM-mode $2$ displays a more divergent structure, reflecting its higher OAM order.

\begin{figure}[htbp]
\centering
%\vspace{-15pt}
\subfigure[OAM-mode $-1$.]{
\begin{minipage}{0.45\linewidth}
\centering
\includegraphics[scale=0.2]{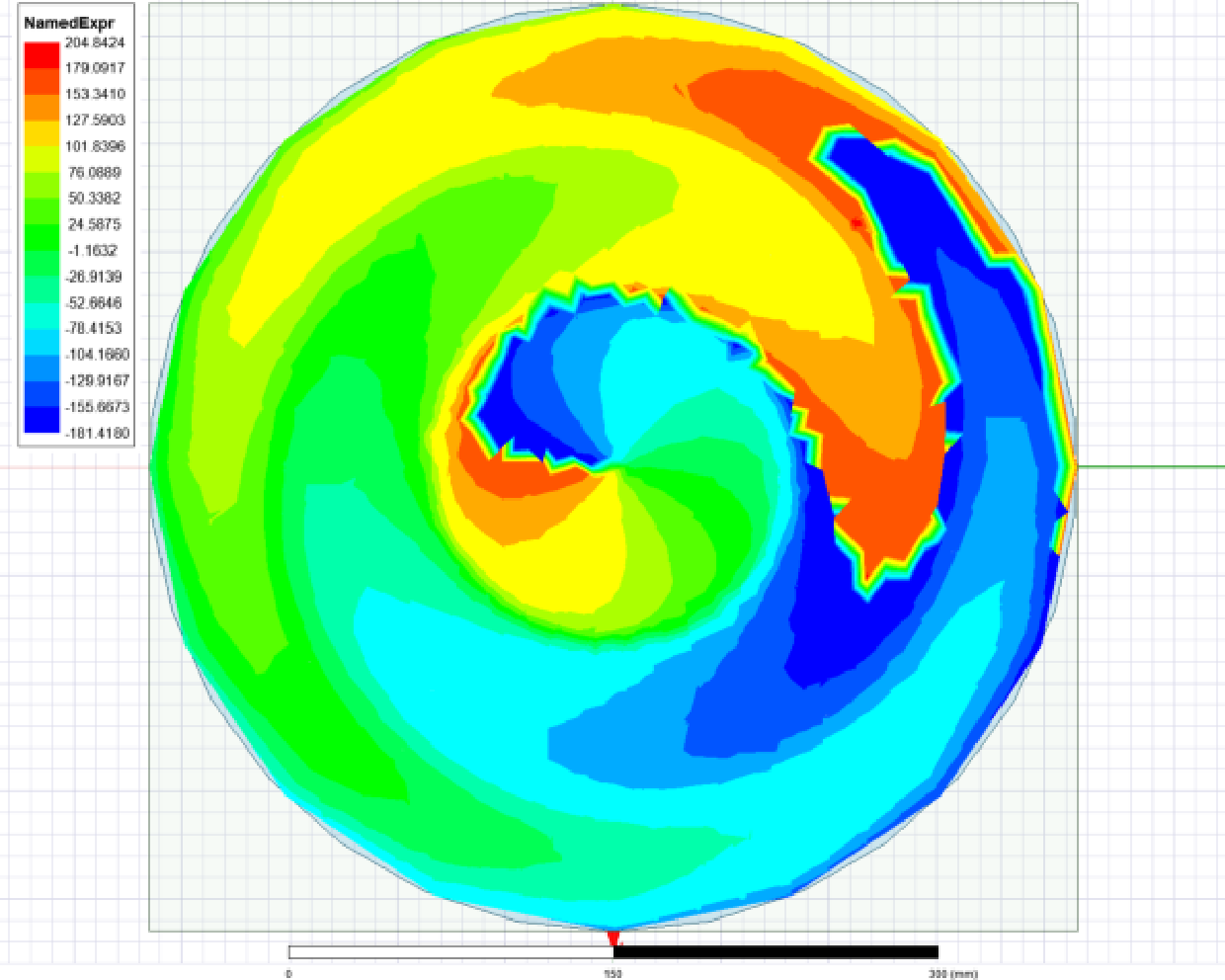}
%\vspace{-30pt}
\end{minipage}
}
\subfigure[OAM-mode $0$.]{
\begin{minipage}{0.45\linewidth}
\centering
\includegraphics[scale=0.2]{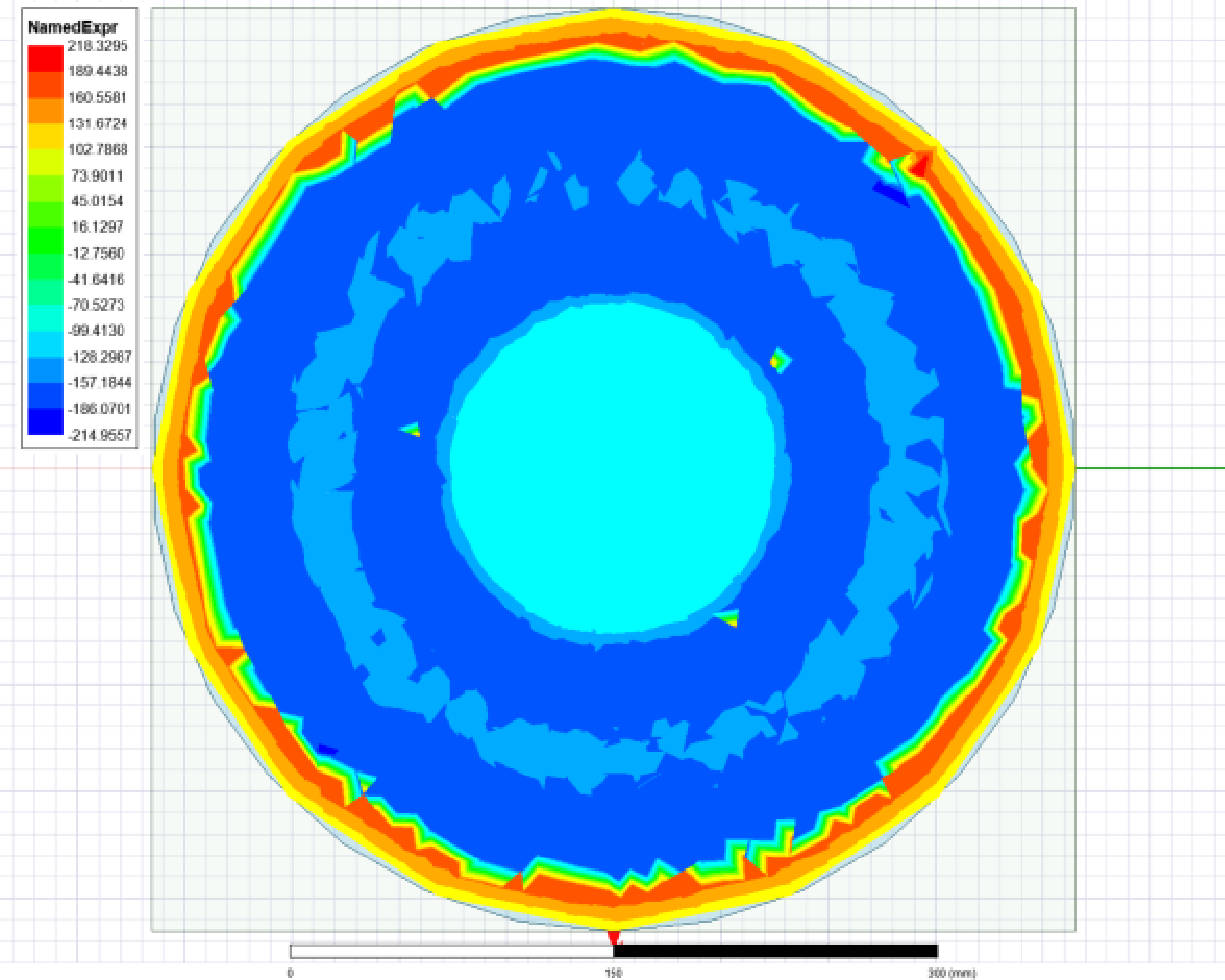}
%\vspace{-30pt}
\end{minipage}
}
\subfigure[OAM-mode $1$.]{
\begin{minipage}{0.45\linewidth}
\centering
\includegraphics[scale=0.2]{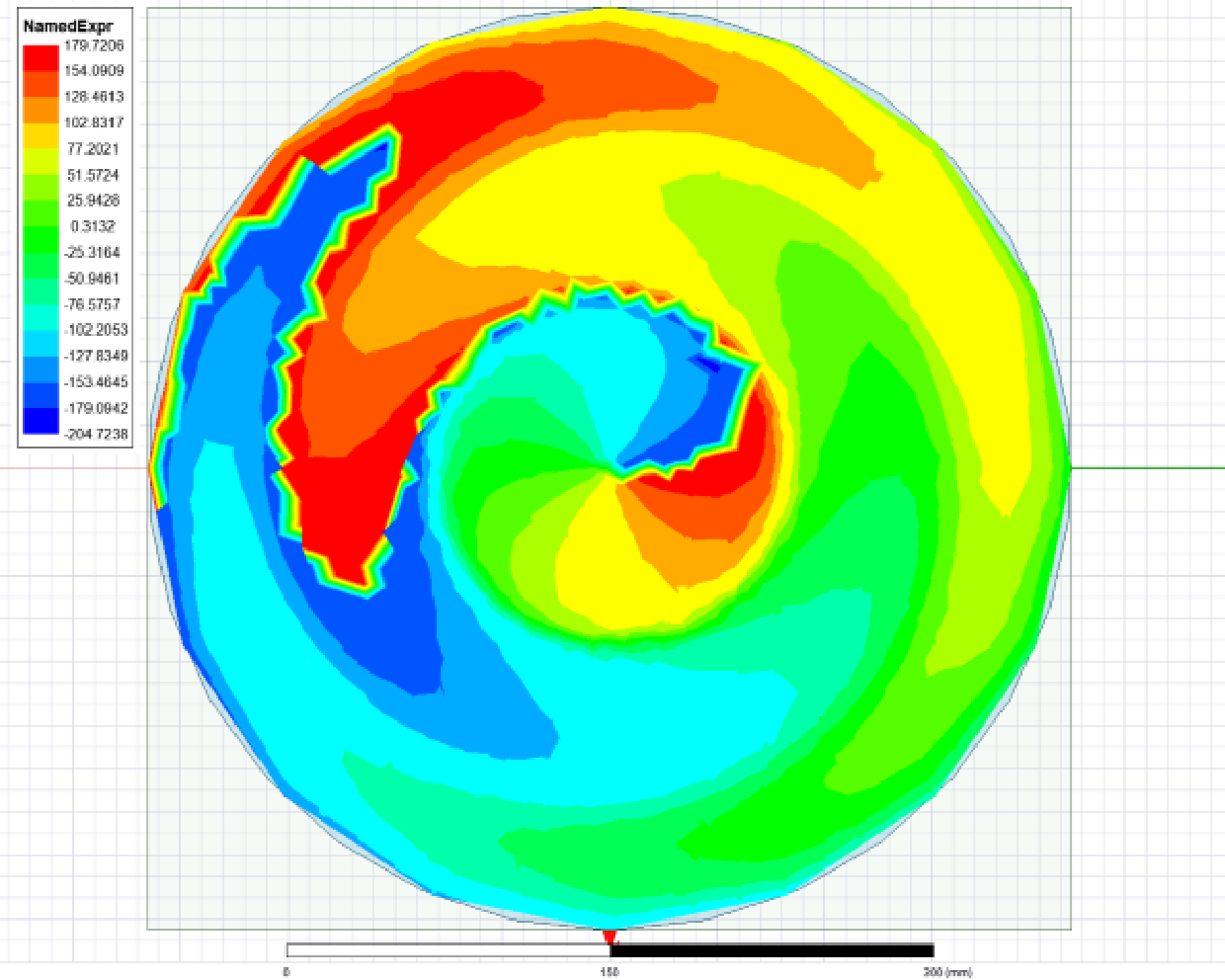}
%\vspace{-30pt}
\end{minipage}
}
\subfigure[OAM-mode $2$.]{
\begin{minipage}{0.45\linewidth}
\centering
\includegraphics[scale=0.2]{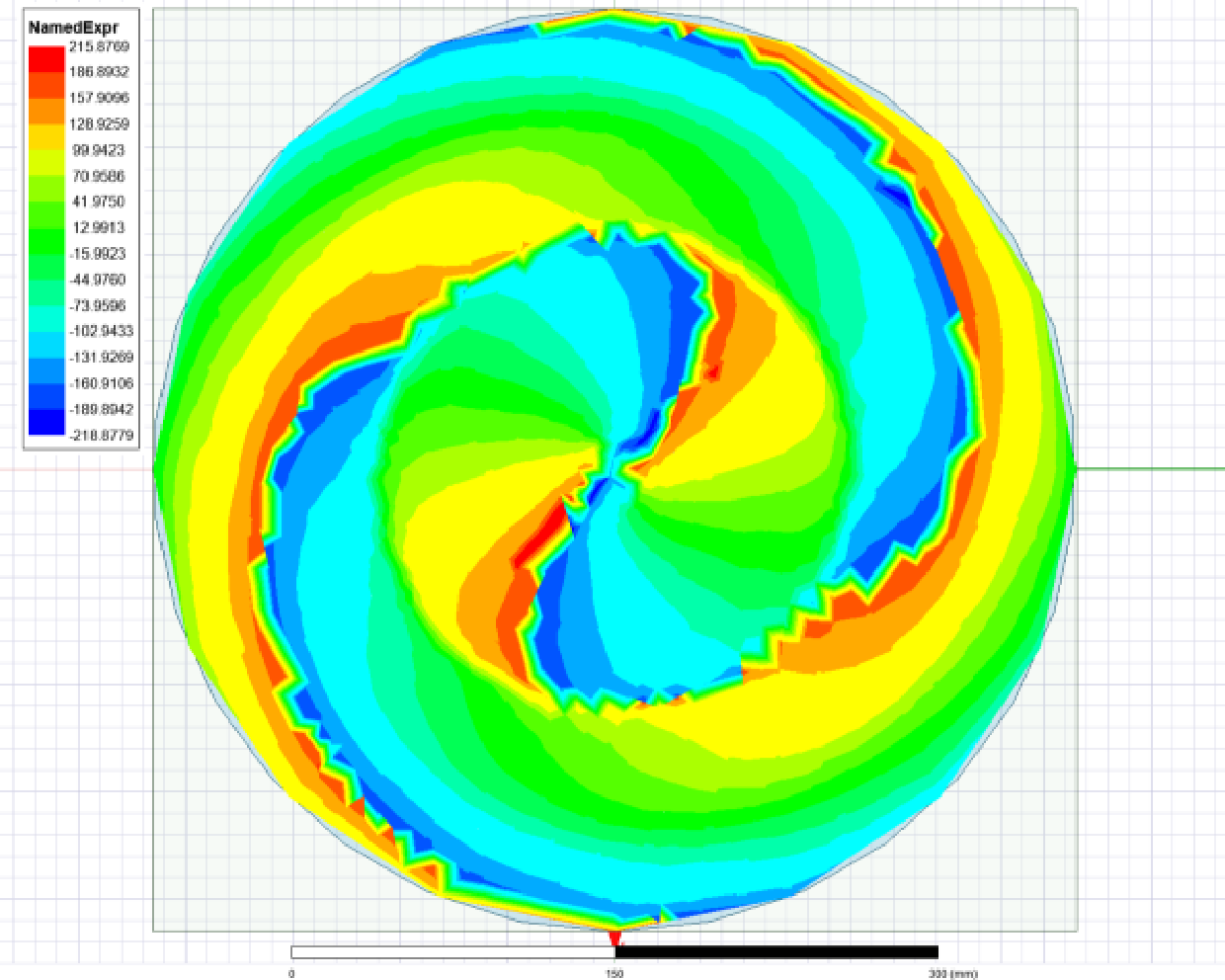}
%\vspace{-30pt}
\end{minipage}
}
\centering
%\vspace{-10pt}
\caption{Phases of $y$-component electric field for OAM-mode $-1$, $0$, $1$, and $2$ simulated by HFSS.} \label{fig:eFiledPhase_HFSS}
%\vspace{-15pt}
\end{figure}
Figure~\ref{fig:eFiledPhase_HFSS} depicts the phases of the $y$-component electric field for OAM modes $-1$, $0$, $1$, and $2$, as simulated by HFSS. The width of the observer plane is also set to $430$ mm. Across all modes, the color gradient represents the phase variation, with the cyclic transition from red to blue denoting different phase values from $-180$ to $180$ degrees. For OAM-mode $0$, the phase distribution appears uniform. The red central region represents a consistent phase across the field, with minor phase changes toward the edges. In contrast, the non-zero OAM modes $-1$, $1$, and $2$ exhibit clear spiral phase patterns, a characteristic of OAM-carrying beams. In OAM-mode $-1$ and OAM-mode $1$, the phase changes smoothly along a helical pattern around the center from $-180$ to $180$ degrees, with $-1$ and $1$ showing opposite rotational directions due to their opposite OAM modes. The spiral pattern becomes more complex in OAM-mode $2$, with two visible phase twists of $720$ degrees, indicating a higher number of phase rotations within the same area. While the OAM mode $0$ shows a constant phase distribution, the higher modes reveal distinctive spiral phase variations, with the complexity increasing as the OAM order rises.

\begin{figure}[htbp]
\centering
%\vspace{-15pt}
\subfigure[OAM-mode $1$ at phase = $0$ deg.]{
\begin{minipage}{0.45\linewidth}
\centering
\includegraphics[scale=0.18]{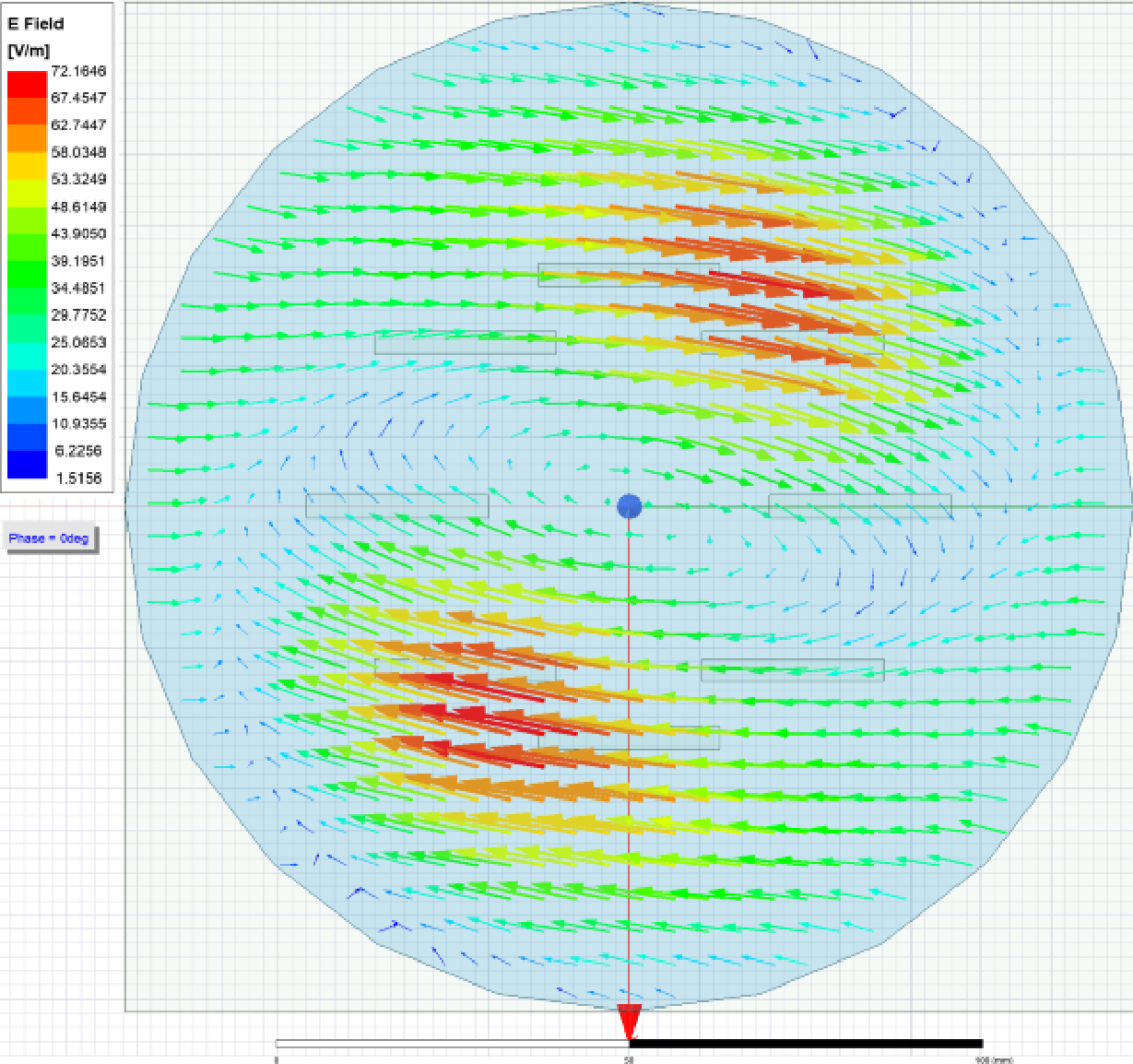}\label{fig:VectorE_Hfss_mode1deg0}
%\vspace{-30pt}
\end{minipage}
}
\subfigure[OAM-mode $1$ at phase = $90$ deg.]{
\begin{minipage}{0.45\linewidth}
\centering
\includegraphics[scale=0.18]{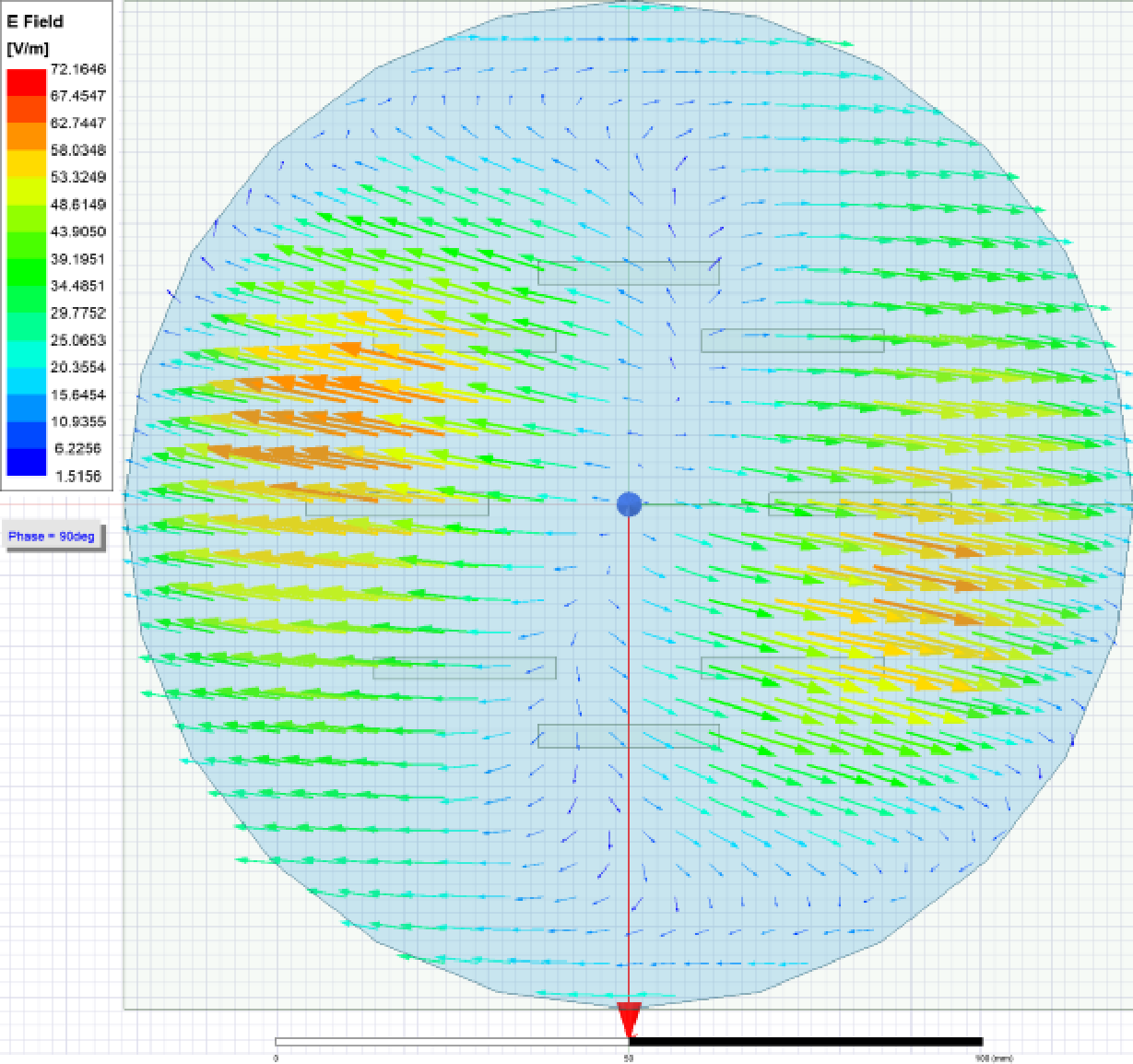}
%\vspace{-30pt}
\end{minipage}
}
\subfigure[OAM-mode $1$ at phase = $180$ deg.]{
\begin{minipage}{0.45\linewidth}
\centering
\includegraphics[scale=0.18]{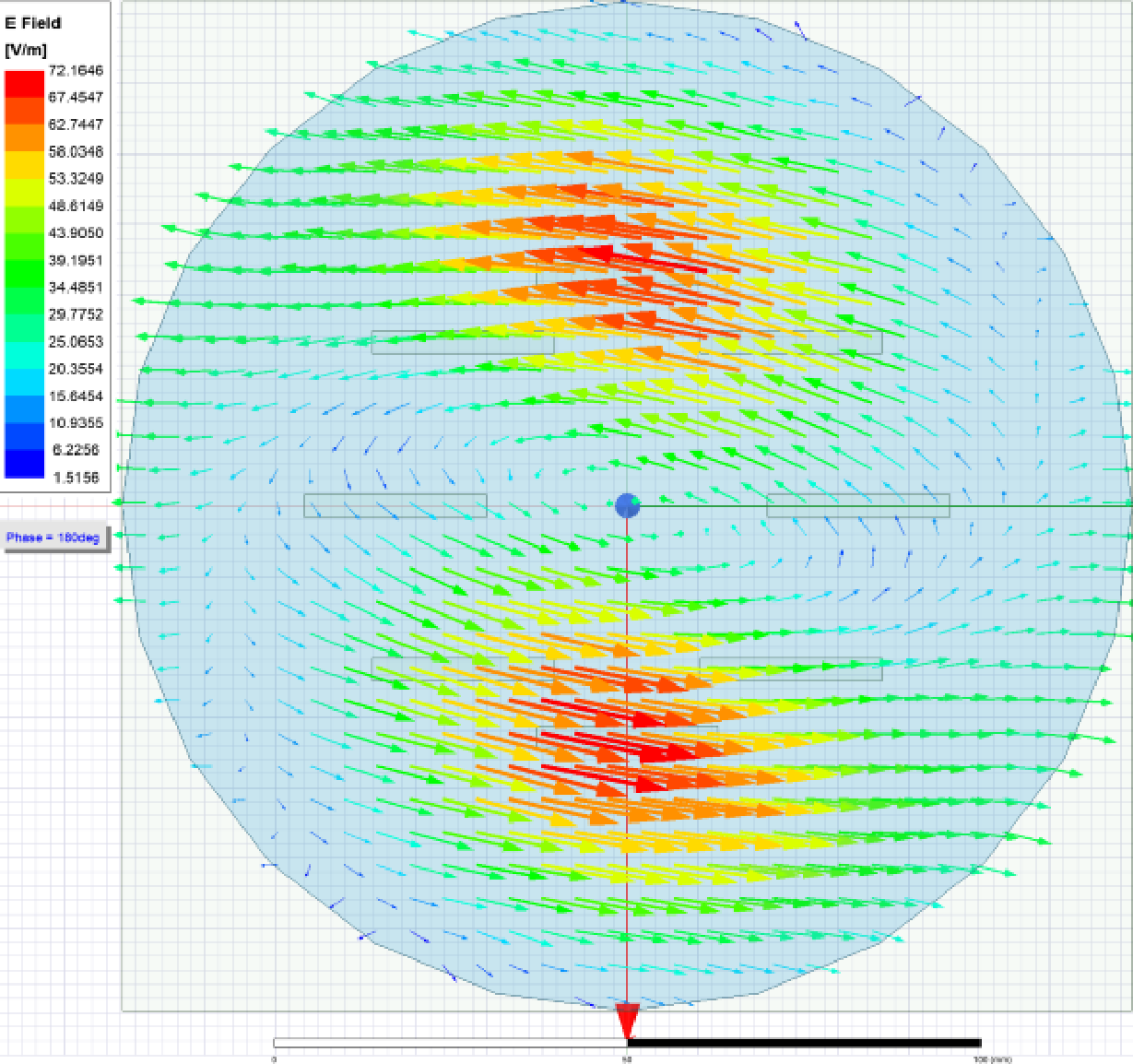}
%\vspace{-30pt}
\end{minipage}
}
\subfigure[OAM-mode $1$ at phase = $270$ deg.]{
\begin{minipage}{0.45\linewidth}
\centering
\includegraphics[scale=0.18]{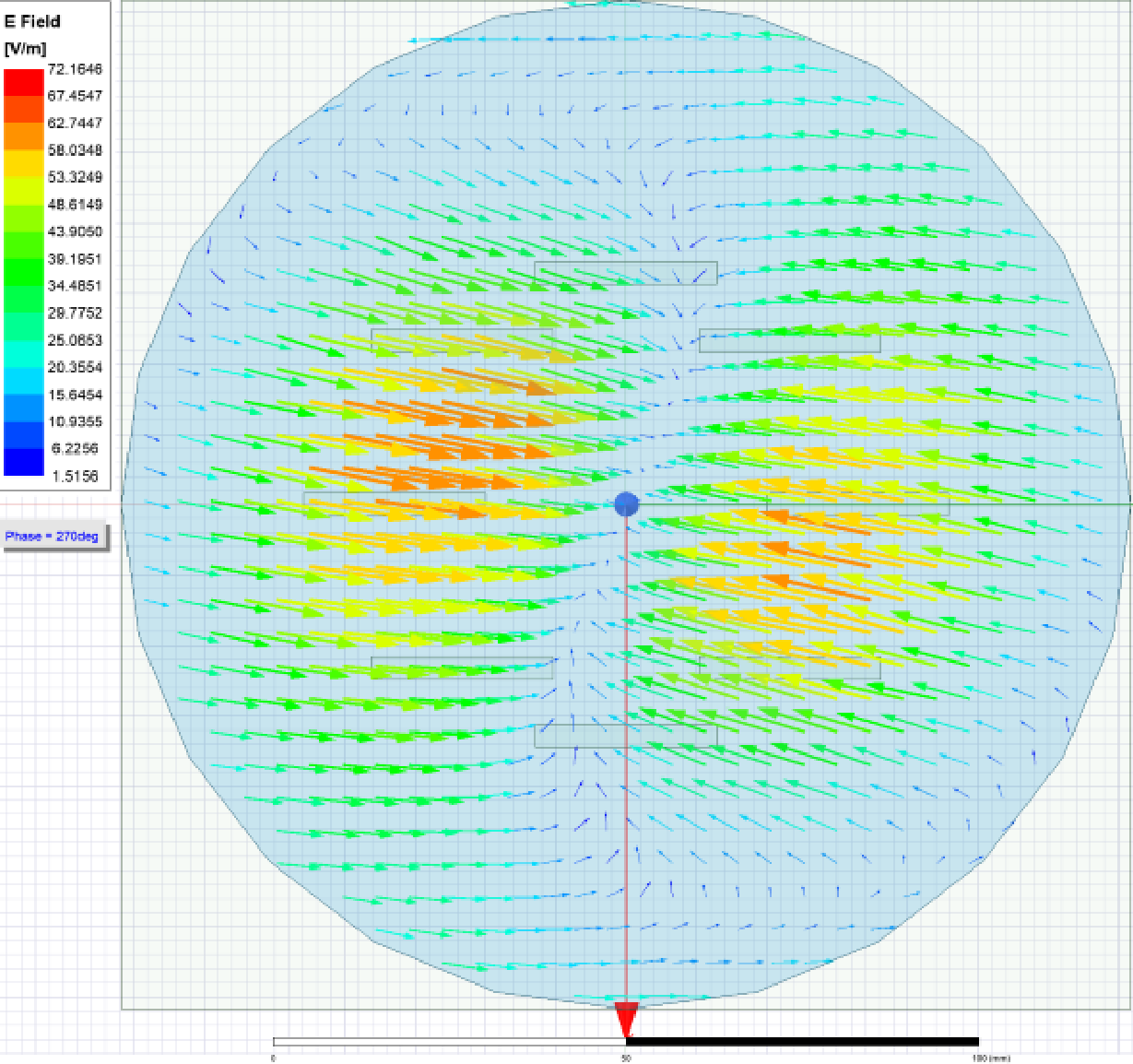}\label{fig:VectorE_Hfss_mode1deg270}
%\vspace{-30pt}
\end{minipage}
}
\subfigure[OAM-mode $-1$ at phase = $0$ deg.]{
\begin{minipage}{0.45\linewidth}
\centering
\includegraphics[scale=0.18]{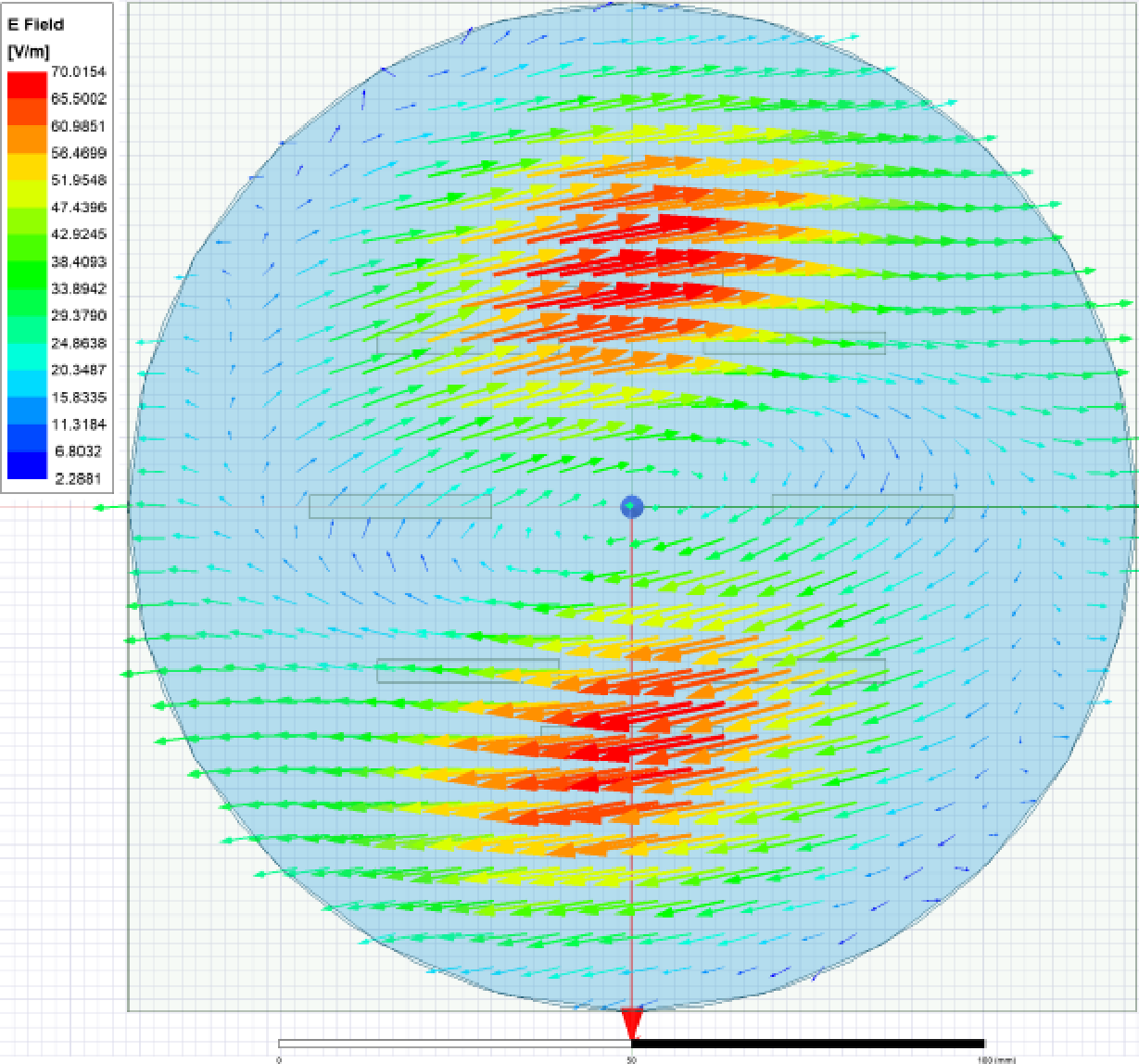}\label{fig:VectorE_Hfss_modeM1deg0}
%\vspace{-30pt}
\end{minipage}
}
\subfigure[OAM-mode $0$ at phase = $0$ deg.]{
\begin{minipage}{0.45\linewidth}
\centering
\includegraphics[scale=0.18]{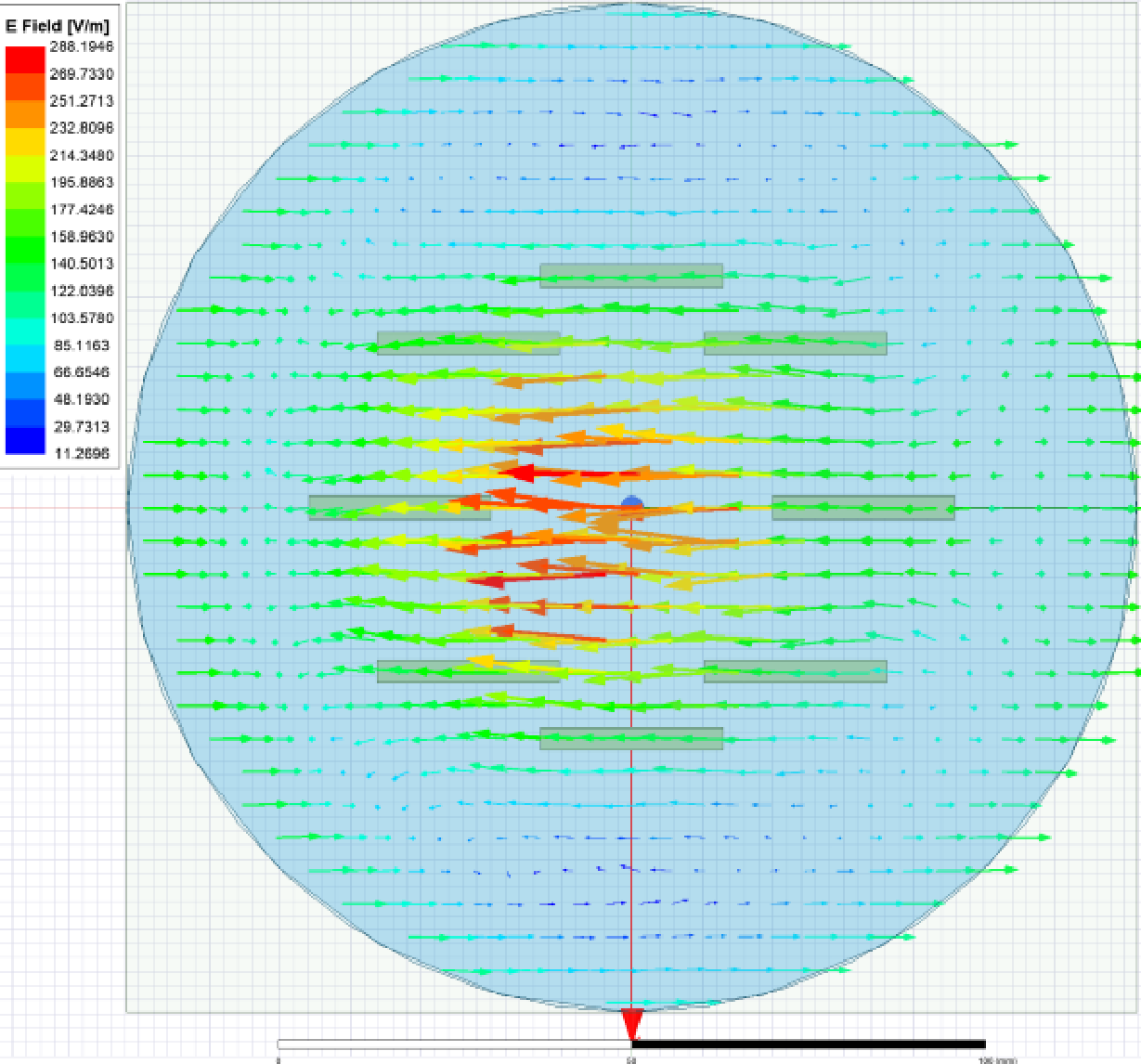}\label{fig:VectorE_Hfss_mode0deg0}
%\vspace{-30pt}
\end{minipage}
}
\centering
%\vspace{-10pt}
\caption{OAM electric fields at different phases simulated by HFSS.} \label{fig:VectorE_Hfss}
%\vspace{-10pt}
\end{figure}
Figure~\ref{fig:VectorE_Hfss} displays the OAM electric fields at different phases as simulated in HFSS. The width of the observer plane is changed to $144$ mm to better show the distribution of the electric field in the center. Figs.~\ref{fig:VectorE_Hfss_mode1deg0} to \ref{fig:VectorE_Hfss_mode1deg270} illustrate the electric fields of OAM-mode $1$ at different phases, while Figs.~\ref{fig:VectorE_Hfss_modeM1deg0} and \ref{fig:VectorE_Hfss_mode0deg0} show the electric fields of OAM-mode $-1$ and $0$, respectively, at a phase of $0$ degrees. For all subplots, the electric fields are primarily oriented along the $y$-axis, consistent with the alignment of the current densities at the source. This is consistent with the numerical results in Fig.~\ref{fig:eFiled}. For OAM-mode $1$ in Figs.~\ref{fig:VectorE_Hfss_mode1deg0} to \ref{fig:VectorE_Hfss_mode1deg270}, the electric field vectors show significant phase-dependent variations. At phase $=0$, the electric field distribution is concentrated in two regions near the center, with the vectors pointing outward. As the phase progresses through $90$, $180$, and $270$ degrees, these concentrated regions rotate around the center, reflecting the helical structure characteristic of OAM modes. The alternating distribution of the high-intensity red regions indicates the phase-dependent rotation of the electromagnetic field. In Figs.~\ref{fig:VectorE_Hfss_modeM1deg0}, OAM-mode $-1$ at phase $=0$ exhibits a similar pattern to OAM-mode $1$, but with an opposite rotational direction. This difference underscores the negative OAM sign of this mode. In Figs.~\ref{fig:VectorE_Hfss_mode0deg0}, the electric field of OAM-mode $0$ shows a more uniform distribution across the circular plane, with less pronounced vector rotation or localized regions of intensity. This aligns with the nature of OAM mode $0$, which lacks OAM, resulting in a more symmetric field structure.

%\begin{figure}[htbp]
%\centering
%%\vspace{-10pt}
%\includegraphics[scale=0.35]{pics//UCA_dual.eps}
%%\vspace{-10pt}
%\caption{Dual UCAs model using lumped ports in HFSS.} \label{fig:UCA_dual}
%%\vspace{-5pt}
%\end{figure}

\begin{figure}[htbp]
\centering
%\vspace{-15pt}
\subfigure[Dual UCAs model using lumped ports in HFSS.]{
\begin{minipage}{1\linewidth}
\centering
\includegraphics[scale=0.32]{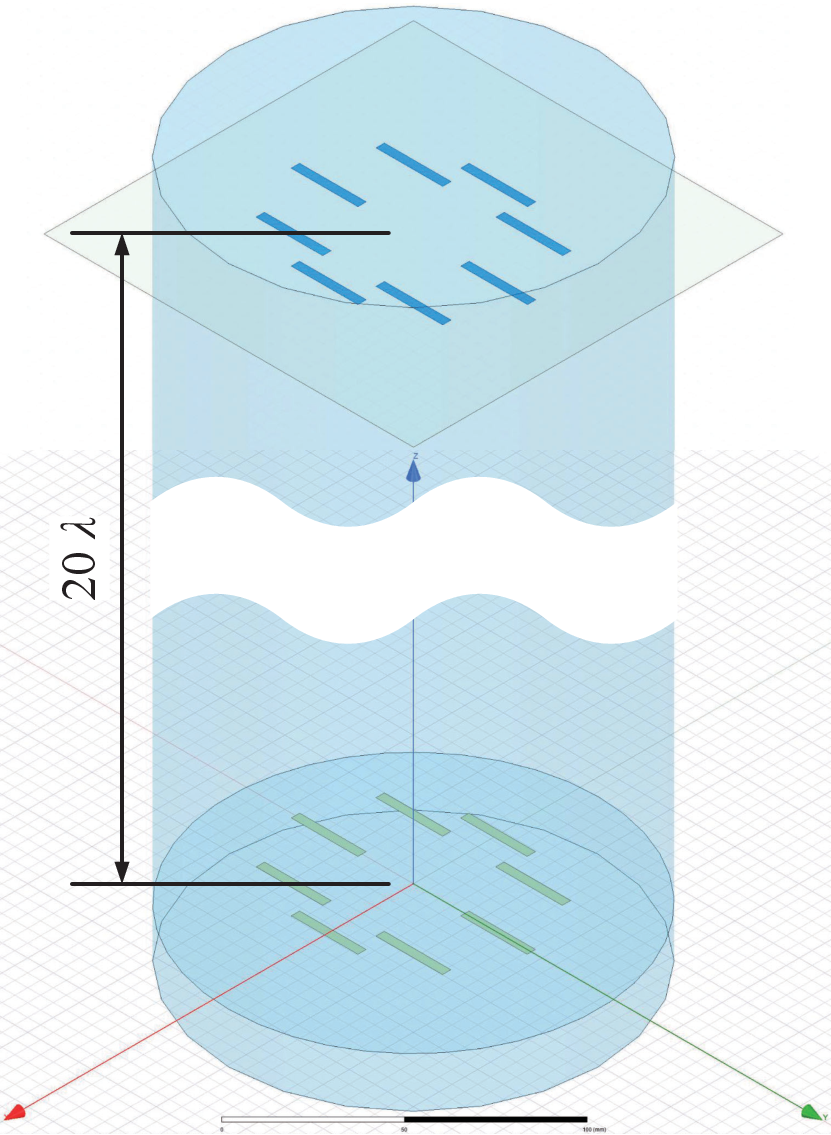}\label{fig:UCA_dual}
%\vspace{-30pt}
\end{minipage}
}
\subfigure[HFSS-simulated channel capacities.]{
\begin{minipage}{1\linewidth}
\centering
\includegraphics[scale=0.55]{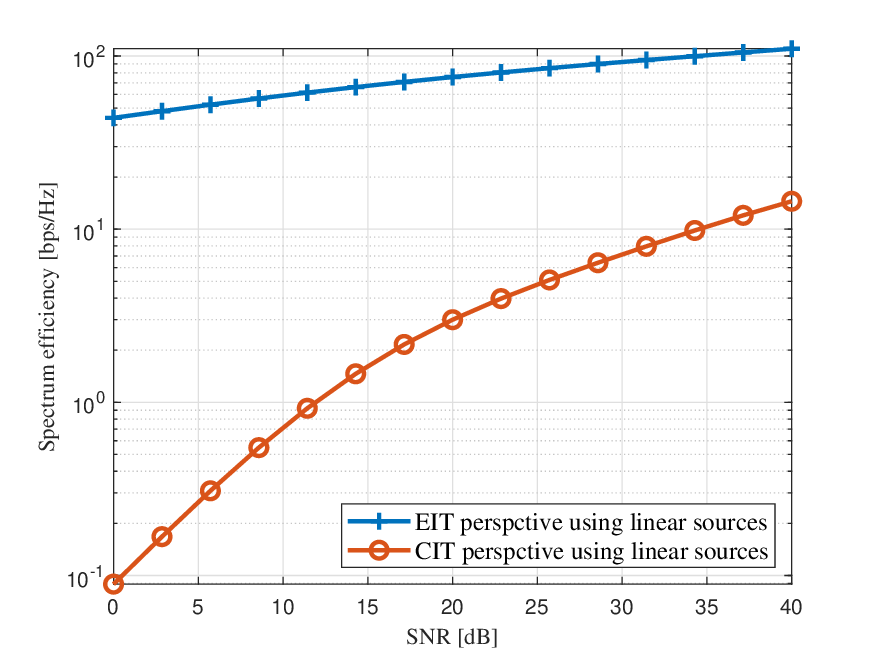}\label{fig:EitOamEnhance_Hfss}
%\vspace{-30pt}
\end{minipage}
}
\centering
%\vspace{-10pt}
\caption{Enhancement of EIT analyzed channel capacity validated by HFSS simulation.}
%\vspace{-10pt}
\end{figure}
To validate the enhancement of analyzing OAM channel capacity through EIT in HFSS simulation, we first present a dual UCA model using lumped ports in Fig.~\ref{fig:UCA_dual} to calculate the channel capacity from the CIT approach for comparison. As shown in in Fig.~\ref{fig:UCA_dual}, the receive UCA is configured with the same parameters as the transmit UCA shown in Fig.~\ref{fig:HfssUCA}. The distance between the transmit and receive UCAs is set to $20\lambda$. All other parameters remain consistent with those in Fig.~\ref{fig:HfssUCA}. The S-parameter matrix for the lumped ports, representing the ratios of power waves between the ports, is derived as follows:
\begin{small}
\begin{align}
\boldsymbol{\mathrm S}=
\begin{bmatrix}
S_{1,1} & S_{1,j} & \cdots & S_{1,N_r+N_t}\\
S_{2,1} & S_{2,j} & \cdots & S_{2,N_r+N_t}\\
\vdots & \vdots & \ddots & \vdots \\
S_{i,1} & S_{i,j} & \cdots & S_{i,N_r+N_t}\\
\vdots & \vdots & \ddots & \vdots \\
S_{N_r+N_t,1} & S_{N_r+N_t,j} & \cdots & S_{N_r+N_t,N_r+N_t}
\end{bmatrix},\label{eq:S_para}
\end{align}
\end{small}
\hspace{-0.15cm}where $S_{i,j}={P_j}/{P_i}$ with $P_i$ and $P_j$ ($1\hspace{-0.1cm}\le\hspace{-0.1cm} i,j\hspace{-0.1cm}\le\hspace{-0.1cm} N_r\hspace{-0.07cm}+\hspace{-0.07cm}N_t$) represent the powers across ports. It is important to note that $\boldsymbol{\mathrm S}$ gives the relationships among all ports, including those between each two transmit ports and the return loss.Thus, the channel matrix, denoted by $\boldsymbol{\mathrm H}_{\rm lumped}$, is derived as the square root of the bottom-left quarter of $\boldsymbol{\mathrm S}$, and is given as follows:
\begin{small}
\begin{align}
\boldsymbol{\mathrm H}_{\rm lumped}=
\begin{bmatrix}
\sqrt{S_{N_t+1,1}}
& \sqrt{S_{N_t+1,2}} & \cdots & \sqrt{S_{N_t+1,N_t}}\\
\sqrt{S_{N_t+2,1}}
& \sqrt{S_{N_t+2,2}} & \cdots & \sqrt{S_{N_t+2,N_t}}
\\ \vdots & \vdots & \ddots & \vdots \\ \sqrt{S_{N_t+N_r,1}}
& \sqrt{S_{N_t+N_r,2}} & \cdots & \sqrt{S_{N_t+N_r,N_t}}
\end{bmatrix}.
\label{eq:H_S_relation}
\end{align}
\end{small}
Therefore, the channel capacity of OAM using linear sources analyzed by CIT, denoted by $C_{\rm CIT}$, can be given as $C_{\rm CIT}={\rm log}_2\left|\boldsymbol{\mathrm I}_{N_r}+\frac{P/N_t}{N_0}{\boldsymbol{\mathrm R}}_{\rm lumped}\right|$, where ${\boldsymbol{\mathrm R}}_{\rm lumped}$ denotes the correlation matrix and is given as ${\boldsymbol{\mathrm R}}_{\rm lumped} = {\boldsymbol{\mathrm W}}^H{\boldsymbol{\mathrm H}}_{\rm lumped}{\boldsymbol{\mathrm W}}\left({\boldsymbol{\mathrm W}}^H{\boldsymbol{\mathrm H}}_{\rm lumped}{\boldsymbol{\mathrm W}}\right)^H$.

%\begin{align}
%C_{\rm CIT}={\rm log}_2\left|\boldsymbol{\mathrm I}_{N_r}+\frac{P/N_t}{N_0}{\boldsymbol{\mathrm R}}_{\rm lumped}\right|,
%\label{eq:C_Hfss}
%\end{align}

%\begin{figure}[htbp]
%\centering
%%\vspace{-10pt}
%\includegraphics[scale=0.55]{pics//EitOamEnhance_Hfss.eps}
%%\vspace{-10pt}
%\caption{Enhancement of EIT-based OAM validated by HFSS simulation.} \label{fig:EitOamEnhance_Hfss}
%%\vspace{-5pt}
%\end{figure}
Figure~\ref{fig:EitOamEnhance_Hfss} illustrates the HFSS-simulated channel capacities for OAM-based wireless communications, calculated using both EIT and CIT approaches, respectively. The OAM channel capacity analyzed by CIT is derived from the S-parameters of the dual UCA model in Fig.~\ref{fig:UCA_dual}. The OAM channel capacity analyzed by EIT is calculated based on the export electric fields of the UCA model shown in Fig.~\ref{fig:HfssUCA}. Specifically, to compute the channel capacity using EIT, signals are transmitted across $8$ OAM modes, ranging from $-3$ to $4$. Then, HFSS is used to export the electric field at $8$ evenly distributed points on a circle with a radius $R_r$, located $20\lambda$ ($1034.5$ mm) away from the transmit UCA, which corresponds to the receive UCA as defined in Section~\ref{sec:systemModel}. Using the DFT transformation of the electric field, the channel matrix and correlation matrix between each transmitting and receiving mode are obtained. The channel capacity is then calculated following Eq.~\eqref{eq:C}. As shown in Fig.~\ref{fig:EitOamEnhance_Hfss}, the EIT method outperforms the CIT method in terms of spectrum efficiency, particularly at lower SNR values. These results are highly consistent with the numerical findings presented in Fig.~\ref{fig:EitOamEnhance}, demonstrating the huge improvement in capacity of analyzing OAM through EIT compared to CIT.

\begin{figure}[htbp]
\centering
%\vspace{-15pt}
\subfigure[UCA model with $N_t = 16$ in HFSS.]{
\begin{minipage}{1\linewidth}
\centering
\includegraphics[scale=0.15]{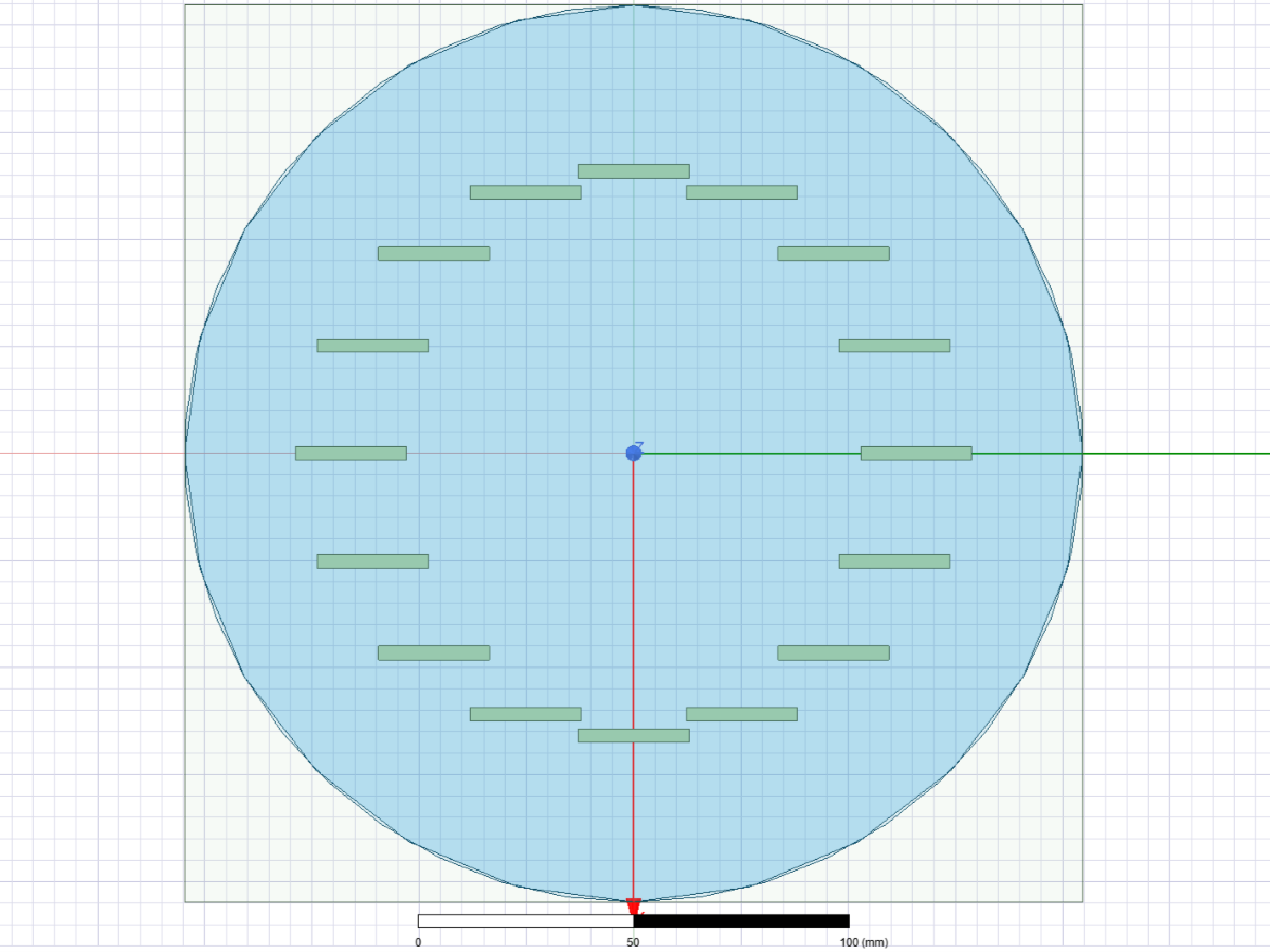}\label{fig:HfssUCA_16}
%\vspace{-30pt}
\end{minipage}
}
\subfigure[Capacities with different numbers of $N_t$ and $N_r$.]{
\begin{minipage}{1\linewidth}
\centering
\includegraphics[scale=0.55]{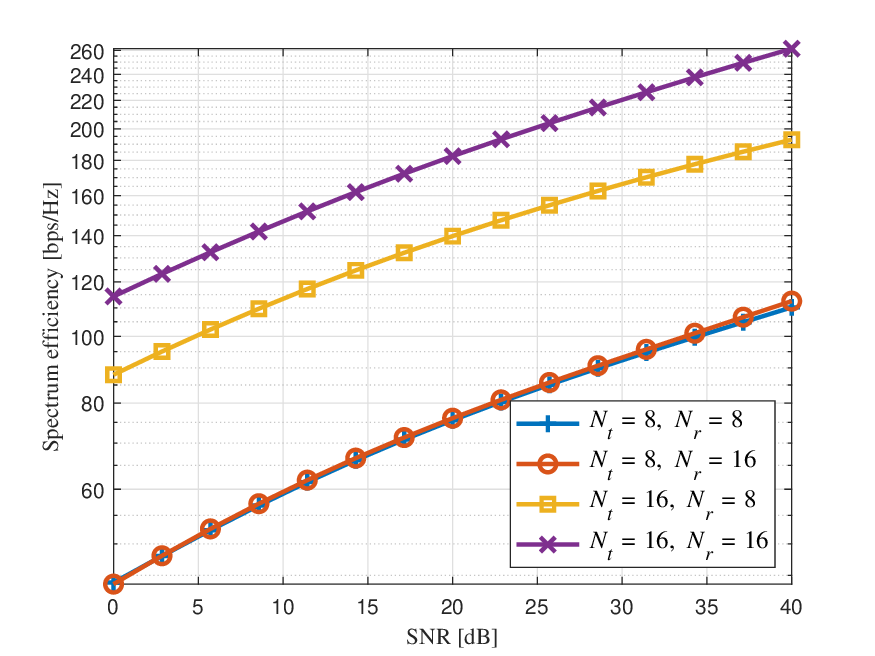}\label{fig:capacityNrNt_Hfss}
%\vspace{-30pt}
\end{minipage}
}
\centering
%\vspace{-10pt}
\caption{HFSS-simulated OAM channel capacities with different numbers of transmitting and receiving mode.}
%\vspace{-10pt}
\end{figure}
Figure~\ref{fig:HfssUCA_16} in the next page presents a transmit UCA model in HFSS equipped with $16$ linear sources and a radius of $4\lambda/\pi$ ($65.86$ mm). Based on the models in Figs.~\ref{fig:HfssUCA} and \ref{fig:HfssUCA_16}, we plot Fig.~\ref{fig:capacityNrNt_Hfss}, illustrating the channel capacities of an OAM-based wireless communication system, where the number of transmitting and receiving modes alternates between $8$ and $16$, and the SNR is set to $20$ dB. All other parameters match those in Fig.~\ref{fig:EitOamEnhance_Hfss}. Consistent with the numerical results in Fig.~\ref{fig:capacityNrNt}, an increase in either transmitting or receiving modes enhances channel capacity in Fig.~\ref{fig:capacityNrNt_Hfss}. When both $N_t$ and $N_r$ are set to $16$, the system achieves significantly higher spectral efficiency compared to configurations with fewer transmitting or receiving modes. The configuration with $N_t = 8$ and $N_r = 8$ yields the lowest capacity. However, when $N_r$ increases to $16$ while keeping $N_t = 8$, capacity improves only slightly. This result differs from Fig.~\ref{fig:capacityNrNt} due to the variation in transmitter-receiver distance. Increasing $N_t$ to $16$ with $N_r = 8$ significantly boosts capacity, as the higher number of transmitting modes enhances orthogonal channel availability. In summary, both $N_t$ and $N_r$ positively impact system capacity, although an increase in $N_t$ alone notably improves capacity, whereas an increase in $N_r$ alone has minimal impact.

\begin{figure}[htbp]
\centering
%\vspace{-10pt}
\includegraphics[scale=0.55]{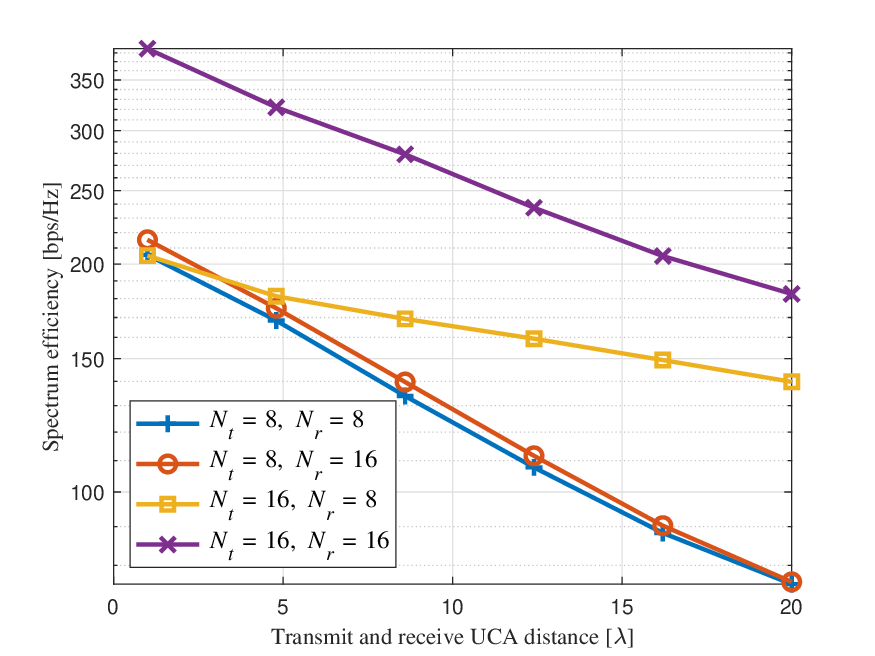}
%\vspace{-10pt}
\caption{HFSS-simulated OAM channel capacities with different transmit and receive UCA distances.} \label{fig:capacityDistanceNrNt_Hfss}
%\vspace{-5pt}
\end{figure}
Figure~\ref{fig:capacityDistanceNrNt_Hfss} depicts the OAM channel capacities for varying distances between the transmit and receive UCAs, ranging from $1\lambda$ ($51.7$ mm) to $20\lambda$ ($1034.5$ mm), across four configurations of transmitting and receiving modes: $(N_t = 8, N_r = 8)$, $(N_t = 8, N_r = 16)$, $(N_t = 16, N_r = 8)$, and $(N_t = 16, N_r = 16)$. All other parameters remain consistent with those in Fig.~\ref{fig:capacityNrNt_Hfss}. In Fig.~\ref{fig:capacityDistanceNrNt_Hfss}, the channel capacities decrease for all configurations as the transmit-receive UCA distance increases, aligning with the numerical results in Fig.~\ref{fig:capacityDistanceNrNt}. The configuration with both $N_t = 16$ and $N_r = 16$ consistently achieves the highest spectral efficiency across all distances. Configurations with $N_t = 8$ exhibit lower spectral efficiencies and experience more pronounced capacity declines as the distance increases, eventually converging at longer distances. The configuration where $N_t = 16$ and $N_r = 8$ shows intermediate performance, with capacities higher than the $N_t = 8$ cases but not as high as when both $N_t$ and $N_r$ are set to $16$.

\begin{figure}[htbp]
\centering
%\vspace{-15pt}
\subfigure[Transmit UCA of different radii.]{
\begin{minipage}{1\linewidth}
\centering
\includegraphics[scale=0.27]{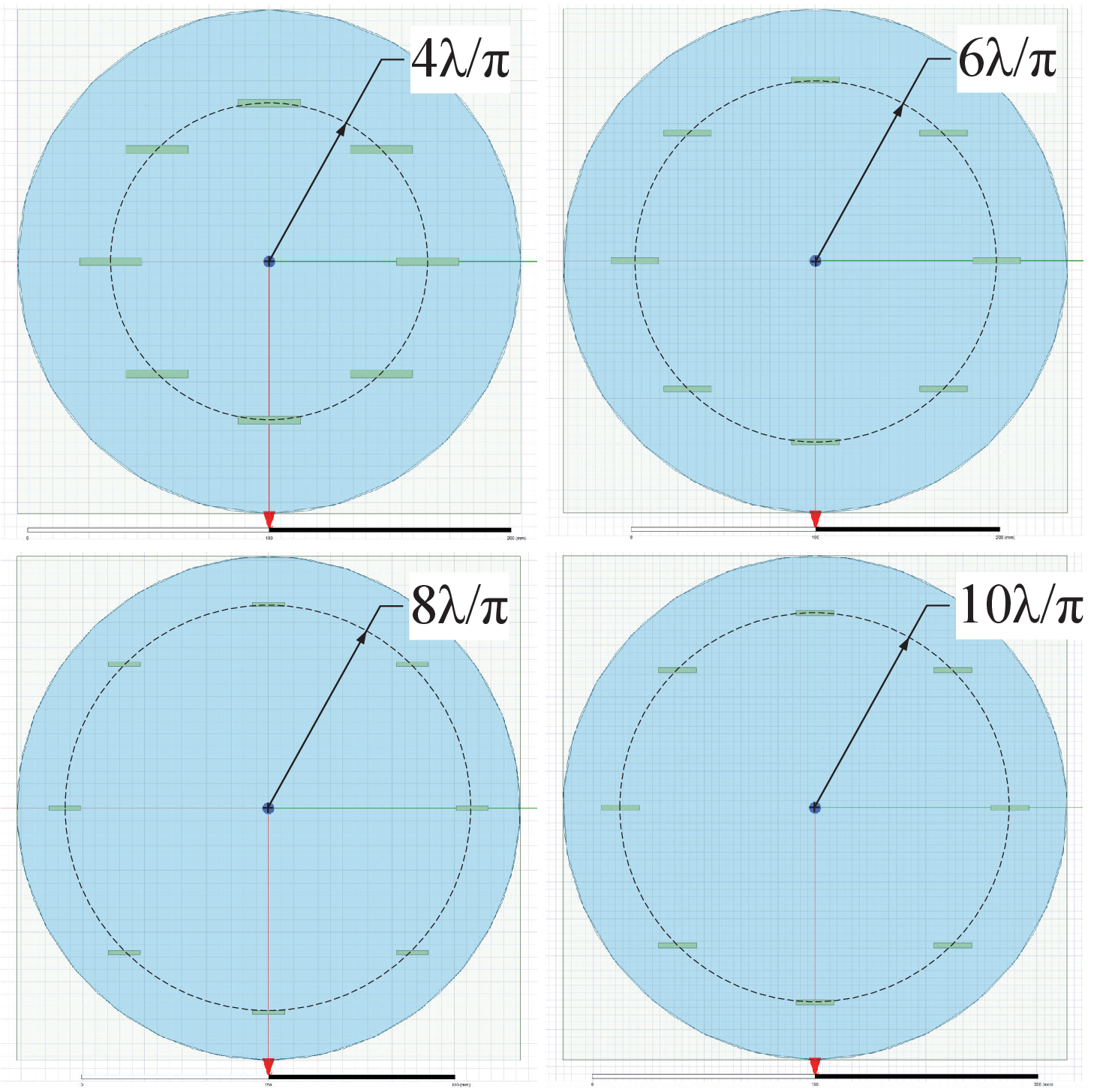}\label{fig:UCA_Rt}
%\vspace{-30pt}
\end{minipage}
}
\subfigure[Transmit UCA radius impact.]{
\begin{minipage}{1\linewidth}
\centering
\includegraphics[scale=0.55]{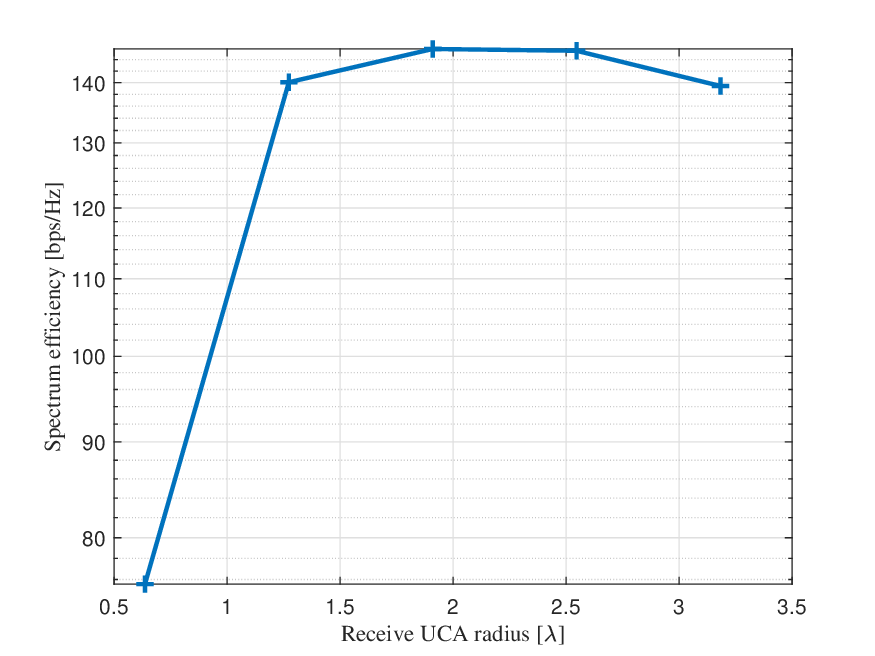}\label{fig:capacityRt_Hfss}
%\vspace{-30pt}
\end{minipage}
}
\centering
%\vspace{-10pt}
\caption{HFSS-simulated OAM channel capacities with different transmit UCA radii.}
%\vspace{-10pt}
\end{figure}
To analyze the impact of the transmit UCA radius on OAM channel capacities, we model four UCAs with different transmit UCA radii, ranging from $4\lambda/\pi$ ($65.9$ mm) to $10\lambda/\pi$ ($164.6$ mm), as shown in Fig.~\ref{fig:UCA_Rt}. Based on results from the models in Figs.~\ref{fig:HfssUCA} and \ref{fig:UCA_Rt}, Fig.~\ref{fig:capacityRt_Hfss} illustrates the effect of varying transmit UCA radii on the channel capacity of the OAM-based wireless communication system. As depicted in Fig.~\ref{fig:capacityRt_Hfss}, an increase in the transmit UCA radius leads to a noticeable improvement in channel capacity, with a sharp rise in spectrum efficiency as the radius grows from $2\lambda/\pi$ ($32.9$ mm) to $4\lambda/\pi$ ($65.9$ mm), consistent with the numerical results. However, after peaking between $6\lambda/\pi$ ($98.8$ mm) and $8\lambda/\pi$ ($131.7$ mm), further increases in radius result in a slight decrease or leveling off in spectrum efficiency.

\begin{figure}[htbp]
\centering
%\vspace{-10pt}
\includegraphics[scale=0.55]{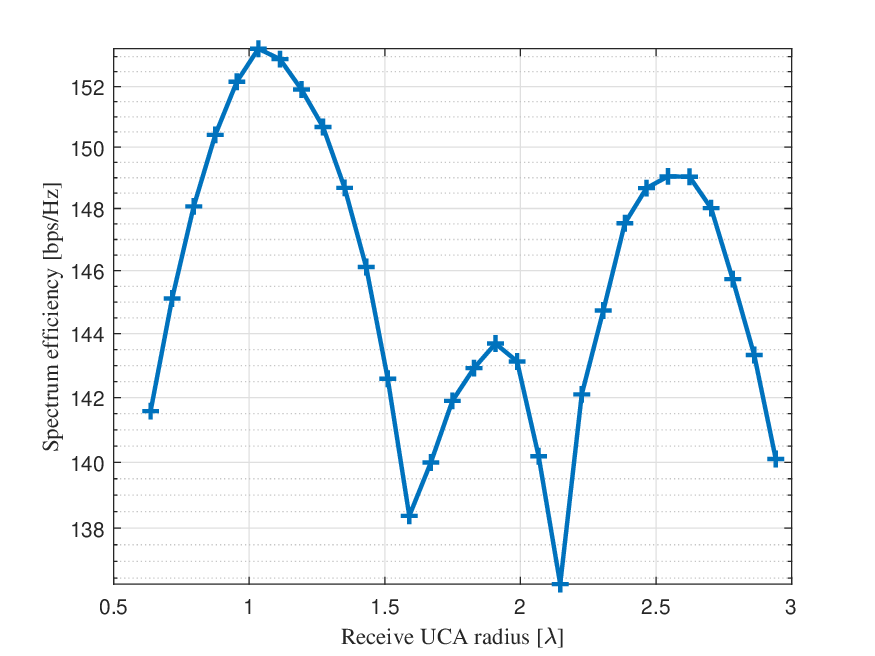}
%\vspace{-10pt}
\caption{HFSS-simulated OAM channel capacities with different receive UCA radii.}\label{fig:capacityRr_Hfss}
%\vspace{-5pt}
\end{figure}
Figure~\ref{fig:capacityRr_Hfss} illustrates how the OAM channel capacity varies with receive UCA radius. Unlike the numerical results shown in Fig.~\ref{fig:capacityRrNrNt}, Fig.~\ref{fig:capacityRr_Hfss} reveals a non-monotonic relationship between receive UCA radius and spectrum efficiency, with multiple peaks and valleys. This fluctuation suggests that certain receive UCA radii enhance channel capacity while others may reduce performance. The variation occurs because each OAM mode has a unique maximum power radius and several sub-high power radii. These results highlight the sensitivity of the OAM channel to changes in receive UCA radius and underscore the importance of optimizing the receive UCA radius to maximize system performance.

%The numerical results in Fig.~\ref{fig:capacityRrNrNt}, where the capacity monotonically increases as the receive UCA radius increases,

\section{Conclusions}\label{sec:Conclusion}
In this paper, we analyzed OAM-based wireless communications from an EIT perspective because of its advantages of integrating electromagnetic theory and CIT. The induced electric field of OAM signal was derived by using Green's function. We then derived the channel capacity for OAM-based wireless communication by presenting the autocorrelation and cross-correlation functions of the received OAM signals. Numerical and simulation results validated the channel capacity enhancement via exploration under EIT framework. Furthermore, we analyzed the impact of various parameters on the channel capacity, including numbers of transmitting and receiving mode, transmit and receive UCA distance, and UCA radius. These findings provide valuable insights into the performance and capacity benefits of OAM-based systems analyzed through EIT.

%\begin{appendices}
%\section{Proof for Theorem~\ref{the:channel_matrix}}\label{pro:multi-coil channel}
%
%\end{appendices}

\bibliographystyle{IEEEtran}
\bibliography{References}

\ifCLASSOPTIONcaptionsoff
  \newpage
\fi

\end{document}